\documentclass[twocolumn,superscriptaddress]{revtex4-1}

\usepackage{array}
\usepackage{graphicx}
\usepackage{verbatim}
\usepackage{color}
\usepackage{amsmath}
\usepackage{booktabs}
\usepackage{multirow}
\usepackage{fancyhdr}

\newcommand{\ket}[1]{\left\vert{#1}\right\rangle}

\begin{document}
%\SetWatermarkText{\textsf{DRAFT}}
%\SetWatermarkScale{5}

\title{Design and Analysis of Communication Protocols for Quantum Repeater Networks}
\author{Cody Jones}
\affiliation{HRL Laboratories, LLC, 3011 Malibu Canyon Road, Malibu, California 90265, USA}
\author{Danny Kim}
\affiliation{HRL Laboratories, LLC, 3011 Malibu Canyon Road, Malibu, California 90265, USA}
\author{Matthew T. Rakher}
\affiliation{HRL Laboratories, LLC, 3011 Malibu Canyon Road, Malibu, California 90265, USA}
\author{Paul G. Kwiat}
\affiliation{Department of Physics, University of Illinois at Urbana-Champaign, Urbana, Illinois 61801-3080, USA}
\author{Thaddeus D. Ladd}
\email{tdladd@hrl.com}
\affiliation{HRL Laboratories, LLC, 3011 Malibu Canyon Road, Malibu, California 90265, USA}

\begin{abstract}
We analyze how the performance of a quantum-repeater network depends on the protocol employed to distribute entanglement, and we find that the choice of repeater-to-repeater link protocol has a profound impact on communication rate as a function of hardware parameters.  We develop numerical simulations of quantum networks using different protocols, where the repeater hardware is modeled in terms of key performance parameters, such as photon generation rate and collection efficiency.  These parameters are motivated by recent experimental demonstrations in quantum dots, trapped ions, and nitrogen-vacancy centers in diamond.  We find that a quantum-dot repeater with the newest protocol (``MidpointSource'') delivers the highest communication rate when there is low probability of establishing entanglement per transmission, and in some cases the rate is orders of magnitude higher than other schemes.  Our simulation tools can be used to evaluate communication protocols as part of designing a large-scale quantum network.  
\end{abstract}

\maketitle
\thispagestyle{fancy}

\section{Introduction}
Quantum information technology applies quantum-mechanical effects to implement beyond-classical applications.  For example, quantum key distribution (QKD) provides the tamper-evident establishment of a secure private key~\cite{BB84,Ekert1991,Bennett1992,Gisin2007}.  In QKD, fundamental properties of quantum mechanics prevent an eavesdropper from intercepting the transmission of a secret key without revealing their interference to authenticated parties.  The unique capabilities of quantum-secure communication and other applications~\cite{Gisin2007,Barz2012,VanMeter2014} motivate research into developing quantum networks~\cite{Briegel1998,Duan2001,Gisin2007,Kimble2008,Sangouard2011,Jouguet2013,Nauerth2013}.

As with classical networks, quantum networks must address engineering concerns like synchronization and latency, though there are further challenges for storage and transmission of quantum information.  Quantum communication depends on the faithful transmission of quantum bits (qubits), which are inherently fragile.  Famous results like the no-cloning theorem~\cite{Wootters1982} exclude the possibility of copying or ``amplifying'' quantum signals, as one could do with classical signals.  Instead, quantum communication that is robust to loss and error can be achieved by using quantum repeaters~\cite{Briegel1998,Simon2007,Yuan2008,Jiang2009,Munro2010}.  Quantum repeaters can transmit, store, and perform logic on qubits, and these operations allow repeaters to herald successfully transmitted signals and to ``distill'' purified quantum information states using error correction~\cite{Gisin1995,Bennett1996_distill,Bennett1996_distill2,Kwiat2000,Zhao2003,Jiang2009,VanMeter2009,Fowler2010,Munro2012,Muralidharan2014}.  These error-suppression protocols provide robustness against both imperfect transmission and the meddling of an eavesdropper.

The purpose and scope of this paper are as follows.   We focus on designing the communication protocol between two neighboring repeater nodes in order to maximize network performance.  To make our analysis concrete, we assume that repeaters are connected with optical fiber, though free-space transmission is a simple extension of our methods.  Furthermore, we focus on quantum technologies that couple controllable quantum memory with single photons and transfer entanglement through two-photon interference, such as trapped ions~\cite{Simon2003,Moehring2007}, diamond nitrogen-vacancy centers~\cite{Childress2006,Dolde2013,Bernien2013}, and quantum dots~\cite{Gao2012,DeGreve2012,Schaibley2013}.  We explain our reasoning for targeting these technologies in Section~\ref{sec::preliminaries}; succinctly, managing photon loss is crucial for designing quantum networks, and the interference and detection of two single-photon signals enables reliable determination of whether a signal was received or lost in transmission while also being more robust to path-length fluctuations than single-detection schemes~\cite{Simon2003,Chen2007}.

The paper begins with some preliminary considerations for distributing entanglement in Section~\ref{sec::preliminaries}.  Section~\ref{sec::link_design} examines three protocols for establishing entanglement between repeaters, and we simulate the performance of these protocols in Section~\ref{sec::simulation} using hardware parameters representative of recent experimental work.  Section~\ref{sec::conclusion} summarizes our results and discusses related communication schemes that we chose not to examine, though they are appropriate for future work.

\section{Preliminaries}
\label{sec::preliminaries}
\begin{figure}
\centering
  \includegraphics[width=8.3cm]{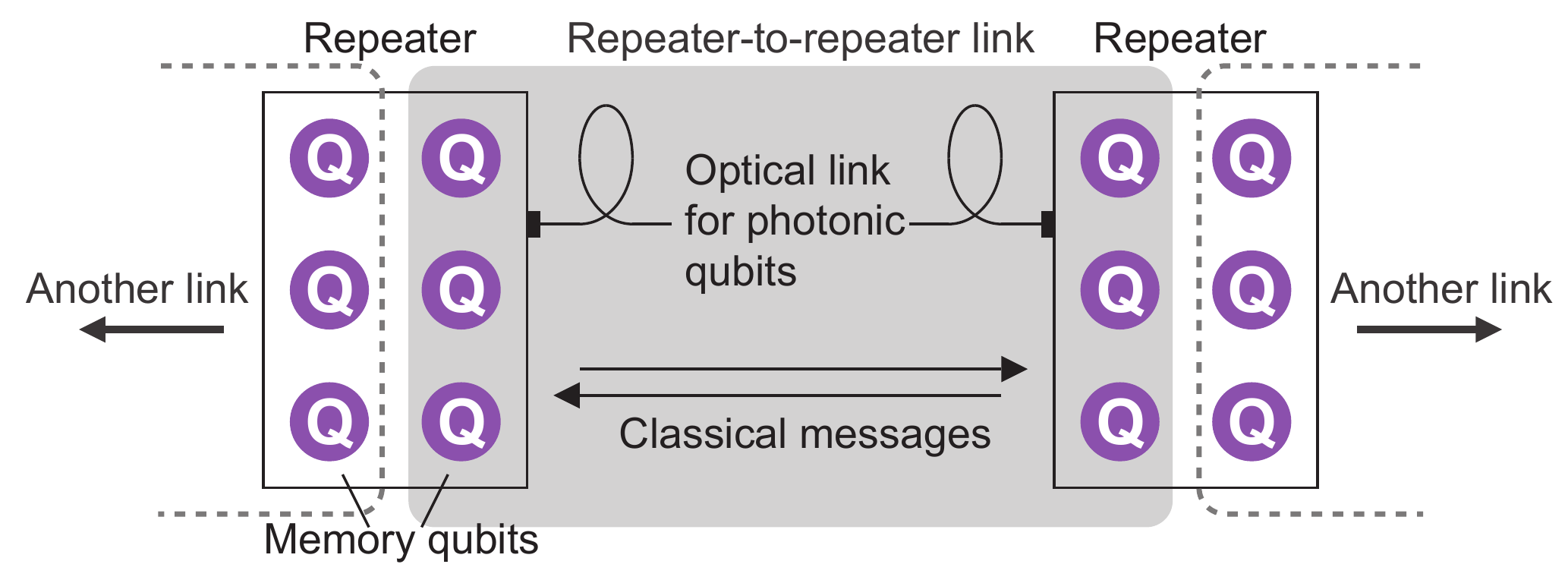}
  \caption{Schematic of a link between two quantum repeaters, showing the basic components.  Every repeater interfaces with two or more links, and the repeater apportions qubits to each of its links.  In this case, two repeaters share a link (gray box), and they participate in links facing left and right (dashed lines) with other repeaters not shown.}
  \label{fig::repeater_general}
\end{figure}

We begin by listing a few features common to any of the repeaters we consider.  As shown in Fig.~\ref{fig::repeater_general}, each repeater has some number of controllable memory qubits that must have long coherence times (of order 10~ms) and low-error gates to act on these memory qubits (error per gate below 0.1\%).  The memory qubits may be protected with error correction~\cite{VanMeter2009,Jiang2009,Fowler2010,Munro2012,Muralidharan2014} to extend their coherence time or suppress gate error.  Furthermore, there is an interface for generating an entangled state between a memory qubit and a single-photonic qubit.  Since there is no ambiguity in this paper, we will simply say photon to mean a photonic qubit.  For example, memory/photon entangled states have been demonstrated for ions, nitrogen-vacancy centers in diamond, and quantum dots~\cite{Moehring2007,Dolde2013,Bernien2013,Gao2012,DeGreve2012,Schaibley2013}.  The memory-photon entanglement is a resource for generating entanglement between repeaters, by swapping entanglement using the photons (described below).  

Each quantum repeater in a network has multiple optical links to its neighbors.  In this paper, we focus on protocols for efficiently distributing entanglement across the link between two repeaters.  As in Fig.~\ref{fig::repeater_general}, each repeater will devote some of its memory qubits to each active link.  Repeaters establish entangled qubit pairs with multiple neighbors to mediate network-wide entanglement~\cite{VanMeter2014}.

Designing a protocol to distribute entanglement in a quantum network is a non-trivial engineering problem, and several fundamental challenges must be addressed.  Signals between two repeater nodes can only travel at the speed of light, but the two repeaters need to make coordinated, synchronous actions.  We take the speed of light in fiber to be $c/n$, where $n \approx 1.5$ is the index of refraction for silica fiber.  We assume that both quantum and classical signals propagate at this speed (ignoring any other networking delays, for simplicity).  Two quantities for delay times are important to analyzing these protocols.  The first is $\tau_{\mathrm{link}} = nL/c$, which is the communication delay between two repeaters separated by link distance $L$.  The second is $\tau_{\mathrm{clock}}$, which is the minimum time needed to either reset a memory qubit or allow detectors to recover from a prior detection event.  The value of $\tau_{\mathrm{clock}}$ depends on both the employed hardware and the choice of protocol.

Each repeater requires a clearly defined protocol for managing its resources.  As a guiding principle, we seek to minimize the amount of time that memory qubits are ``locked up'' while the information needed for the next action is unavailable, and delays from multiple back-and-forth communications should be avoided wherever possible.  To realize the highest communication rate, the state-machine protocol local to each node needs to infer what the repeater at the other end of the link is doing, and what information it has available.  For some events, an immediate action is executed (such as processing or erasing a memory qubit), while in other cases the protocol waits for more information.  For simplicity, we assume that these protocols are synchronized by a distributed clock.

A defining feature of the protocols we consider is that they distribute entanglement using single-photon qubits and two-photon interference~\cite{Duan2003,Feng2003,Simon2003,Moehring2007,Chen2007,Yuan2008}.  Specifically, we perform Bell-state measurement in an apparatus known as a Bell-state analyzer (BSA)~\cite{Michler1996,Luetkenhaus1999,Duan2003,Feng2003,Simon2003,Chen2007,Yuan2008}.  The type of BSA used in our schemes employs linear optics and single-photon detectors, meaning that it necessarily succeeds for at most 50\% of attempts~\cite{Calsamiglia2001}.  We implement a BSA that can reliably measure two of four Bell states, which is sometimes called a ``partial BSA.''  A successful Bell-state measurement is indicated by two coincident single-photon detection events, so detectors with high efficiency and low dark count rate are critical to the entanglement-distribution schemes we consider.

The Bell-state measurement performs entanglement swapping~\cite{Zukowski1993,Pan1998,Duan2003,Feng2003,Simon2003,Chen2007,Yuan2008}, so that if the interfering photons were entangled to memory qubits in two repeaters, the memory qubits are projected into an entangled state on successful Bell-state measurement.  When this happens, classical messages are sent to both repeaters.  We design protocols that assume either zero or one photons will enter the BSA at each input port.  If two photons enter one port and lead to detection events marked as successful, this is an error that degrades fidelity of entanglement.  We assume that the probability a successful BSA outcome results from multiple-photon emission or detector dark counts is sufficiently small (on the order of 1\% or less) to be suppressed through entanglement distillation~\cite{Gisin1995,Bennett1996_distill,Bennett1996_distill2,Kwiat2000,Zhao2003,VanMeter2009}.

Photon loss is the primary concern in our analysis, and two-photon detection enables ``loss heralding,'' where the protocol can reliably determine if both photons arrived at the BSA.  When a photon propagates through fiber over distances of 10~to 100~km, it has substantial probability of being lost due to material absorption or other imperfections.  To address this problem, we exploit the property that a photon is quantized, so it either arrives at its destination or is lost.  In contrast to entanglement schemes employing bright coherent states, a detection apparatus that is expecting a single photon to arrive within a narrow time window can mark the absence of detection as a failed transmission attempt, while a successfully transmitted qubit is stored in quantum memory.  The bit of information indicating whether the qubit was lost is sent to the repeaters with classical communication, allowing the repeaters to coordinate entanglement distribution through the protocols we examine in Section~\ref{sec::link_design}.  

As a final consideration, we assume that all entangled memory qubits are used at the end of one round of entanglement distribution, which we explain more explicitly in the next section.  Simply put, two repeaters act only on information available through their shared link and do not hold qubits in storage while waiting for information to arrive from elsewhere in the network.  While more complex protocols may be required in some applications, this assumption simplifies our analysis by keeping protocols contained to a single link and serves as a good approximation for the implementations of QKD that we simulate.  The performance of each entanglement protocol therefore depends on how frequently entanglement can be attempted and how quickly information confirming entanglement or photon loss is available.

\section{Protocols for Distributing Entanglement}
\label{sec::link_design}
This section considers three different protocols for distributing entanglement between two neighboring repeaters using single photons, which we label as \texttt{MeetInTheMiddle}, \texttt{SenderReceiver}, and \texttt{MidpointSource}.  We chose these labels for conceptual clarity and consistent presentation, though versions of these schemes appear in several proposals in the literature.  To analyze these protocols, we examine both the arrangement of hardware components and the time-dependent behavior of signal transmission, memory management, and distributed decision making.  Implementing a protocol requires control logic in the repeater to respond to new information (classical signals or detector outcomes) as it arrives.  For each of the link protocols, we describe and analyze a sequence of operations to implement the associated communication scheme.

\subsection{Meet-In-The-Middle}
\begin{figure}
\centering
  \includegraphics[width=8.3cm]{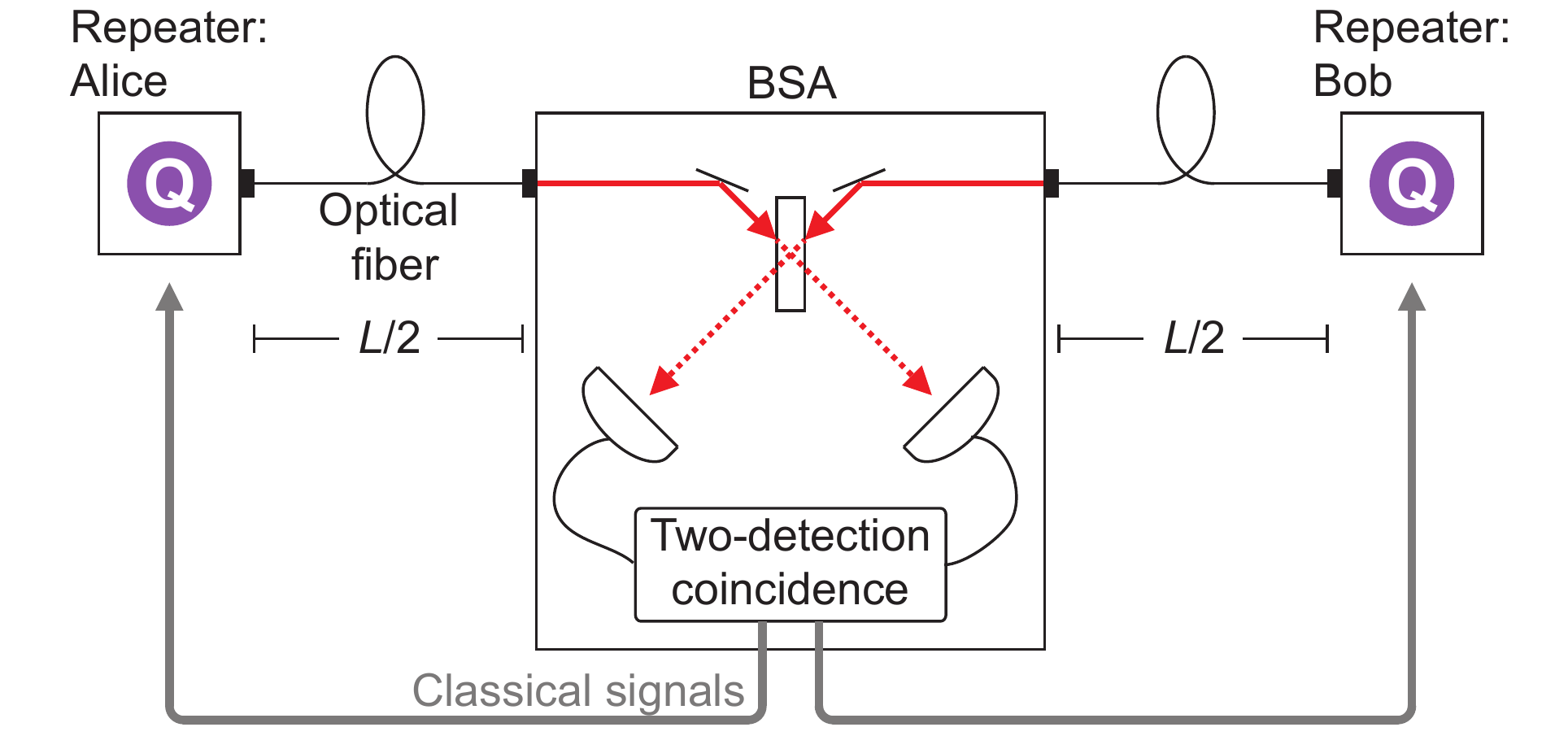}
  \caption{Hardware arrangement for \texttt{MeetInTheMiddle}.  Repeaters Alice and Bob are separated by distance $L$, which is typically tens of kilometers.  Memory/photon pairs are generated simultaneously at each repeater node, and the photons are coupled into optical fiber.  The photons interfere in a BSA located at the channel midpoint, and a successful entanglement swap, which is communicated to both repeaters with classical signals, projects the corresponding memory qubits into an entangled state.  The scheme can be modified to an asymmetric arrangement where the BSA is located at any position in the optical channel.  In this case, the photons are generated at such times that they arrive simultaneously at the BSA.}
  \label{fig::MITM}
\end{figure}

The \texttt{MeetInTheMiddle} protocol has two repeater nodes, at both ends of an optical link, transmit photons to a BSA at the midpoint of the link.  Each photon is entangled with a memory qubit in the sending node.  The arrangement of hardware components for \texttt{MeetInTheMiddle} is shown in Fig.~\ref{fig::MITM}.  When the BSA succeeds in swapping entanglement, the pair of memory qubits, one in each repeater node, is projected into an entangled state.  This protocol was introduced in Refs.~\cite{Duan2003,Feng2003,Simon2003}.

The simplicity of \texttt{MeetInTheMiddle} makes it a good starting point for explaining entanglement-distribution protocols, and elements of \texttt{MeetInTheMiddle} will reappear in the more complex protocols considered later.  As in Figure~\ref{fig::MITM}, the link places repeaters Alice and Bob at distance $L$ apart and a BSA at the midpoint.  In this symmetric arrangement, both repeaters have the same number of memory qubits connected to the link.  Alice and Bob generate memory/photon entangled qubit pairs timed such that the photons interfere in the BSA.  The corresponding memory qubits are projected into an entangled state when the BSA succeeds, though neither Alice nor Bob can use this entanglement until a confirmation signal returns from the BSA after speed-of-light delay of $\tau_{\mathrm{link}}$.  We say that the memory qubit is ``locked up'' during this waiting period.  

The sequence of transmissions is as follows.  Entanglement is attempted in ``rounds,'' where each round has duration
\begin{equation}
\tau_{\mathrm{round}} = \tau_{\mathrm{link}} + N \tau_{\mathrm{clock}},
\label{eqn::MITM_round_time}
\end{equation}
as explained below.  Alice and Bob synchronize the emission of a photon entangled to memory such that Alice's photon and Bob's photon arrive at the BSA at the same time (if they are not lost).  Each repeater has $N$ memory qubits.  If $N > 1$, the repeaters generate photons at regular intervals synchronized to $\tau_{\mathrm{clock}}$, allowing detectors in the BSA to recover if necessary.  This time-division multiplexing is what makes the round duration $\tau_{\mathrm{link}} + N \tau_{\mathrm{clock}}$.  Each attempt at generating entanglement is assigned an identifying number $i \in [1,N]$ for this round, which corresponds to both the photon's position in the communication sequence and the location of the entangled memory qubit in the repeater.  As photons arrive at the channel midpoint, the BSA will transmit a message $\mathcal{M}_i$ to both repeaters containing one of two statements: ``Bell-state measurement succeeded for transmission $i$,'' or the opposite ``did not succeed.''  The control protocol in each repeater processes these messages and determines what next action to take.

\begin{figure}
\centering
  \includegraphics[width=8.3cm]{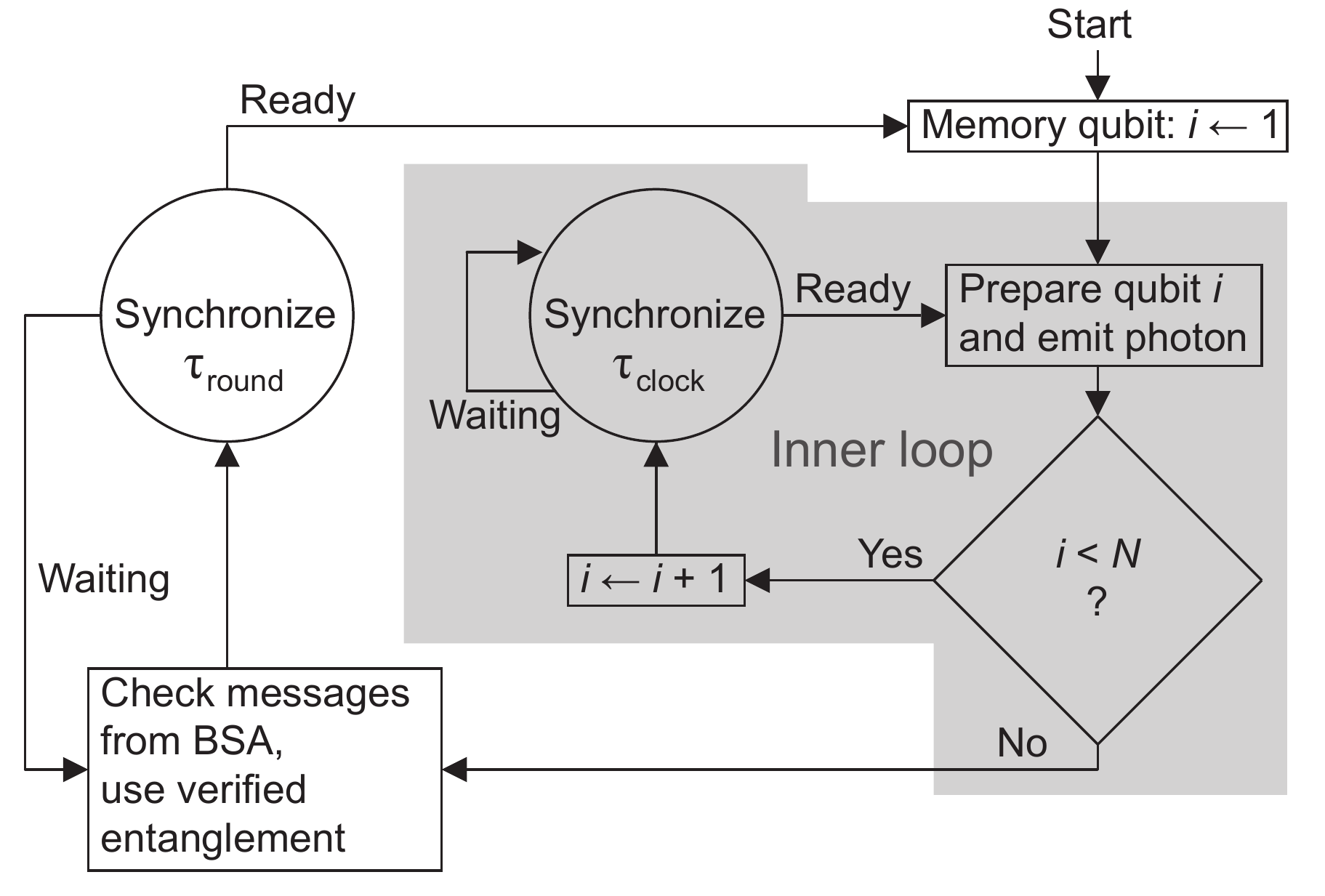}
  \caption{Control protocol for \texttt{MeetInTheMiddle}.  Loop variable $i$ follows the definition in the text.  The statements ``\hbox{$i \leftarrow 1$}'' and ``\hbox{$i \leftarrow i + 1$}'' simply mean ``assign value 1 to $i$'' and ``increment $i$ by 1,'' respectively.  If the repeater has multiple memory qubits ($N > 1$), transmission attempts are synchronized to $\tau_{\mathrm{clock}}$ to allow the BSA to recover.  After using all of the memory qubits to send photons, the protocol waits for responses from the BSA to complete the round.}
  \label{fig::MITM_control}
\end{figure}

The control protocol for \texttt{MeetInTheMiddle} is shown in Fig.~\ref{fig::MITM_control} as a state machine.  To understand this diagram, the symbols have the following meaning: arrows indicate transitions between states; a rectangular box is an action that is executed when the protocol arrives in that state; a diamond is a query for information that branches based on the answer; and a circle is a synchronize query.  The synchronize query enforces clocking behavior, meaning that it will exit through the ``waiting'' branch until an internal clock steps into the next indicated time period (e.g. synchronized to $\tau_{\mathrm{clock}}$ or $\tau_{\mathrm{round}}$).  After receiving the messages $\{\mathcal{M}_i\}$ from the BSA indicating entanglement success or failure, the repeaters use all memory qubits at indices $\{i \vert \mathcal{M}_i \equiv \mathrm{success}\}$.  Having completed the round, the repeaters reset the memory qubits and repeat the process in the next round.  

We have made two assumptions to keep the protocol simple, but these can be relaxed in a more complex protocol for either expanded network functionality or increased performance.  First, we assume that the repeater waits for all messages before using any memory qubits, which is reasonable for $N \tau_{\mathrm{clock}} < \tau_{\mathrm{link}}$.  Otherwise, the first signals confirming entanglement arrive before all memory qubits transmit photons, so there may be an opportunity to reduce memory lock-up time by using memory qubits as soon as entanglement is confirmed.  Second, we assume that all memory qubits are reset at the end of each round, though other protocols may need to preserve entangled memory qubits for multiple rounds.  We leave analysis of these scenarios to future work.

The diagram in Fig.~\ref{fig::MITM_control} is designed to show how efficient \texttt{MeetInTheMiddle} is at distributing entanglement.  Focus on the ``inner loop'' indicated by the grey region.  While the protocol is traversing the inner loop, the repeater is actively attempting entanglement distribution.  After the repeater has filled all of its memory qubits, the ``synchronize $\tau_{\mathrm{round}}$ query''  forces the repeater to wait and check for messages from the BSA.  Let us define a measure of efficiency: the link utilization factor for \texttt{MeetInTheMiddle} is
\begin{equation}
F = \frac{N\tau_{\mathrm{clock}}}{\tau_{\mathrm{round}}},
\end{equation}
which is simply the ratio of time spent in the inner loop to total round time.  If we define $p$ as the probability of successfully projecting two memory qubits into an entangled state, then the average entanglement-distribution rate can be expressed as
\begin{equation}
R = \frac{Np}{\tau_{\mathrm{round}}} = F\frac{p}{\tau_{\mathrm{clock}}}.
\label{eqn::MITM_rate}
\end{equation}
The quantity $R_{\mathrm{ub}} = p/\tau_{\mathrm{clock}}$ is an important quantity, because both $p$ and $\tau_{\mathrm{clock}}$ are fundamental properties of the link distance and hardware.  $R_{\mathrm{ub}}$ is average rate achieved if entanglement is attempted every clock cycle (the protocol is always in its inner loop), which is an upper bound to any achievable rate.  The efficiency of the \texttt{MeetInTheMiddle} protocol is captured by $F$, the fraction of time spent in the inner loop of Fig.~\ref{fig::MITM_control}, because $R = F R_{\mathrm{ub}}$.

\texttt{MeetInTheMiddle} realizes distributed decision making rather easily, but it makes inefficient use of memory qubits because each is locked up for at least $\tau_{\mathrm{link}}$ while waiting on the signal from the BSA, which succeeds with probability $p$.  In an asymmetric scheme where Bob has a shorter distance to the BSA, his qubits are locked up for a shorter duration, and he could potentially use fewer memory qubits, which is the basis of the next protocol we consider.

\subsection{Sender-Receiver}
\label{sec::SR}
The \texttt{SenderReceiver} protocol is similar to \texttt{MeetInTheMiddle}, but the BSA has been moved into a repeater at one endpoint of the optical link.  This rearrangement has interesting consequences for repeater design, and the modification is a precursor to the third protocol studied here, \texttt{MidpointSource}.  \texttt{SenderReceiver} was discussed in Ref.~\cite{Munro2010}.

\begin{figure}
\centering
  \includegraphics[width=8.3cm]{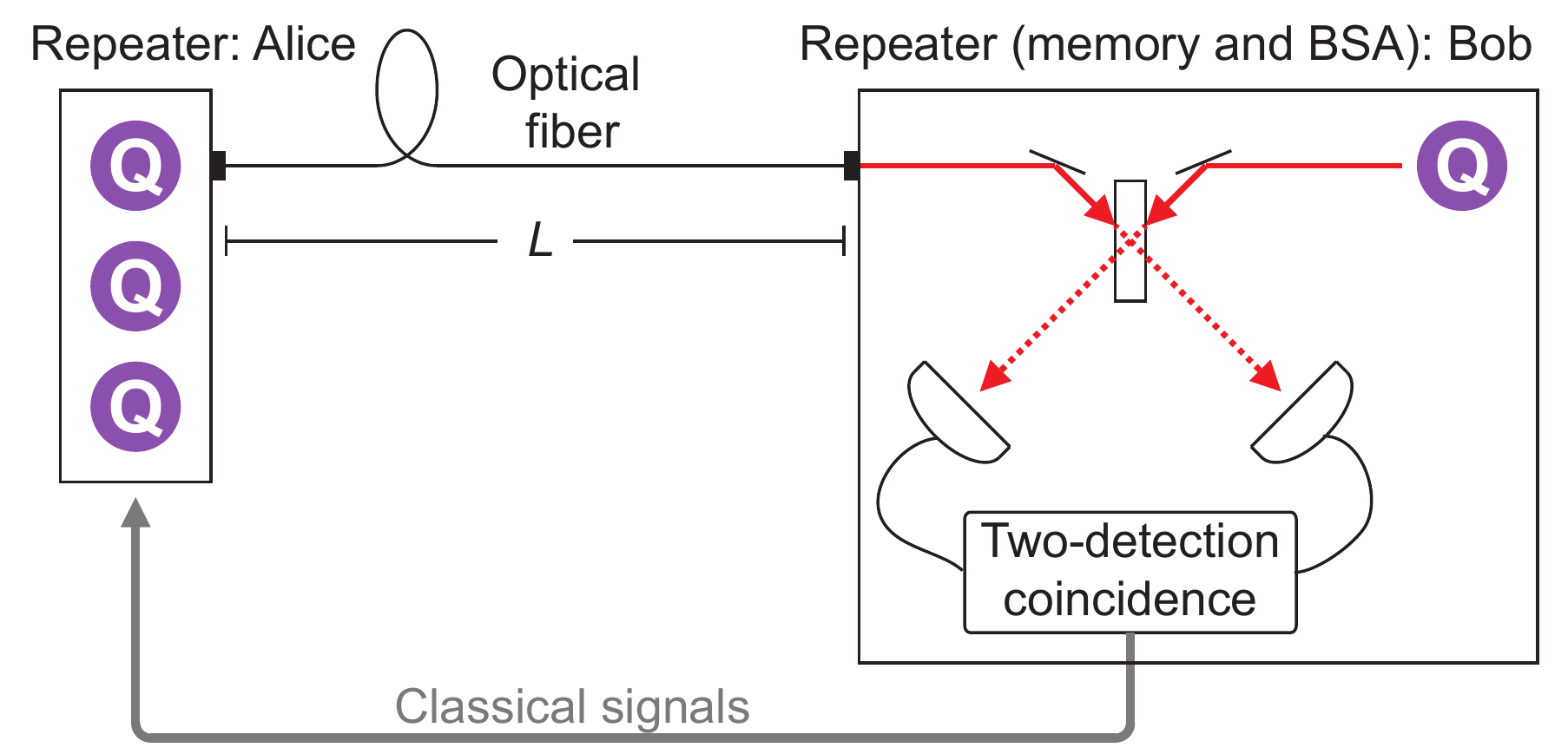}
  \caption{Hardware arrangement for \texttt{SenderReceiver}.  The arrangement is similar to \texttt{MeetInTheMiddle}, but the BSA is now located inside Bob.  As explained in the text, this protocol is most effective when Alice has more memory qubits than Bob (three shown here), because of the asymmetry in delays following BSA interference.  Alice has to wait for round-trip signal propagation to learn if entanglement was established.  Conversely, the BSA is internal to Bob, so he prepares a memory/photon pair just as one of Alice's photons arrives, and he can determine almost instantly whether entanglement was established.}
  \label{fig::SR}
\end{figure}

To understand how \texttt{SenderReceiver} was derived, consider what limits the performance of \texttt{MeetInTheMiddle}.  The bottleneck to communication rate in \texttt{MeetInTheMiddle} is $\tau_{\mathrm{link}}$, the time for a photon to reach the BSA and the return trip for the classical signal that is the result of Bell-state measurement.  When loss heralding is employed, a quantum memory must be locked up while waiting for this result.  \texttt{MeetInTheMiddle} places the BSA in the middle to create symmetric delay for both repeaters, but if the BSA were closer to one of the repeaters, that repeater would be able to make a quicker decision as to whether each local memory qubit was entangled to a qubit in the other repeater.  Taking this concept to its limit, \texttt{SenderReceiver} places the BSA inside Bob's repeater, allowing Bob to make near-instantaneous decisions (limited only by the speed of detectors and control electronics) about whether or not his memory qubit holds useful entanglement.  As the name suggests, we call Alice the sender and Bob the receiver in Fig.~\ref{fig::SR}.

\begin{figure}
\centering
  \includegraphics[width=8.3cm]{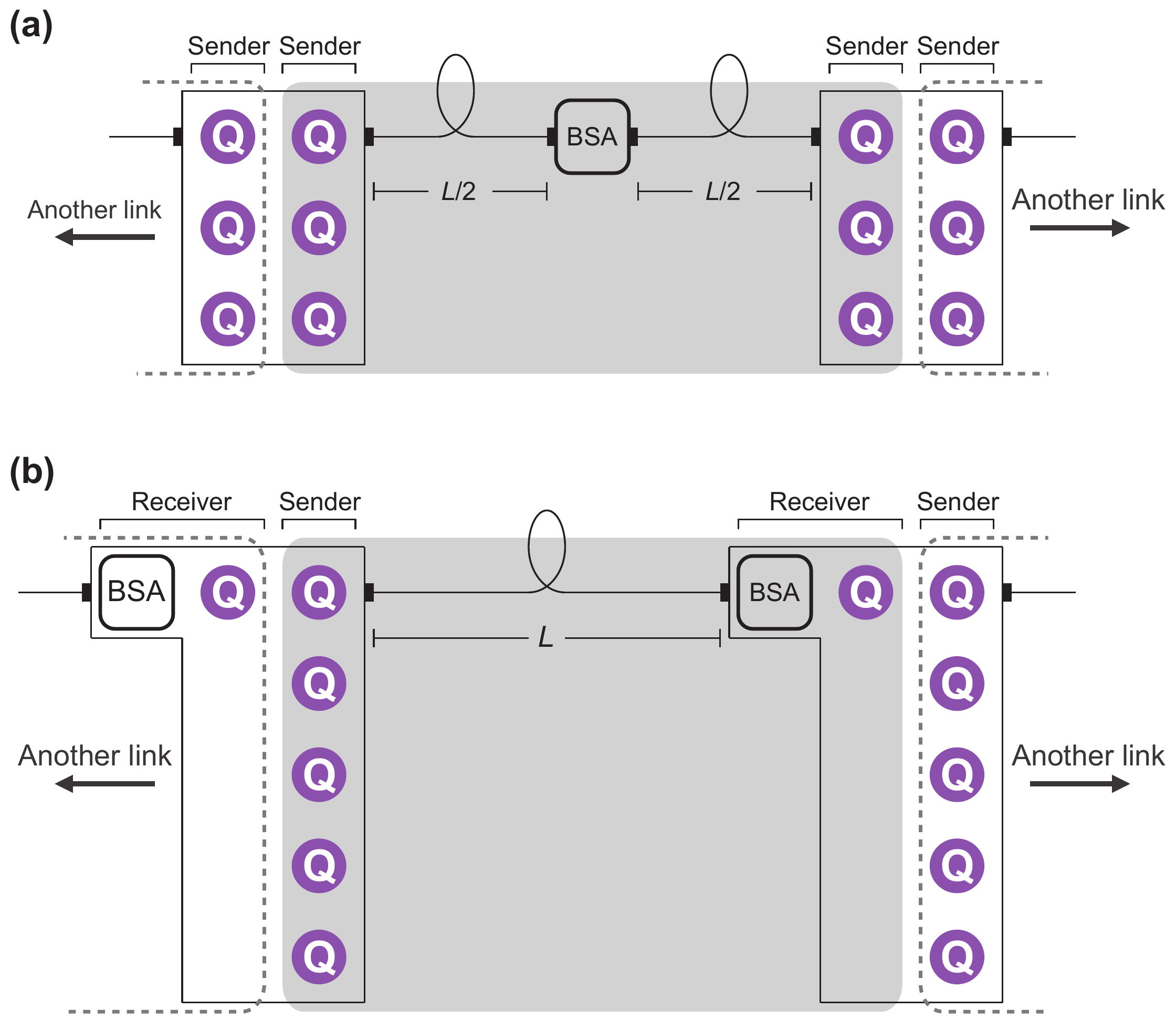}
  \caption{Link interfaces for (a)~\texttt{MeetInTheMiddle} and (b)~\texttt{SenderReceiver}.  In this paper, the type of every link interface is either sender or receiver.}
  \label{fig::sender_and_receiver}
\end{figure}

The ``sender'' or ``receiver'' behavior applies to the interface between memory qubits and photons.  A repeater has at least two interfaces in order to distribute entanglement.  This concept is illustrated in Fig.~\ref{fig::sender_and_receiver}, which compares the link interfaces for \texttt{MeetInTheMiddle} and \texttt{SenderReceiver}.  Whereas \texttt{MeetInTheMiddle} has every link interface act as a sender, \texttt{SenderReceiver} alternates roles between sender and receiver, such that each repeater is a sender in one direction and receiver in the other~\cite{Munro2010}.

The sender operates essentially the same as in \texttt{MeetInTheMiddle}, but the receiver has new behavior.  The receiver works by attempting to ``latch'' an incoming photon from Alice into memory.  Latching here is fundamentally the same as in \texttt{MeetInTheMiddle}, except that Bob can reset his memory immediately on BSA failure.  When a photon arrives, Bob attempts entanglement swapping by producing a memory/photon entangled pair and interfering Alice's photon and his photon in the BSA.  If entanglement swapping succeeds, Alice's photon is transferred into Bob's memory.  We note that, to enable loss heralding, the latching process works indirectly through entanglement swapping using the BSA; Alice's photon is not directly absorbed into a memory qubit.  A key feature of the \texttt{SenderReceiver} protocol is that loss heralding occurs inside the receiver, indicating immediately to that repeater whether latching was successful.

Knowing almost instantly when entanglement is established changes how the repeater operates.  Suppose that Alice has $N_A \gg 1$ memory qubits, allowing her to send many transmissions to Bob, who has fewer memory qubits $N_B < N_A$.  By knowing immediately whether entanglement was established, Bob can process each incoming photon sequentially with the same memory qubit by resetting his qubit after failing to latch Alice's photon.  The inter-transmission time $\tau_{\mathrm{clock}}$ for \texttt{SenderReceiver} is the maximum of the times for detector recovery and memory reset.  If $\tau_{\mathrm{clock}} \ll \tau_{\mathrm{link}}$, then Bob can use fewer memory qubits than Alice because Bob can update his memory based on the latching outcome almost instantly.  However, the memory qubits in Alice are locked up for total round time 
\begin{equation}
\tau_{\mathrm{round}}' = 2\tau_{\mathrm{link}} + N_A \tau_{\mathrm{clock}},
\end{equation}
where we use the prime ($'$) to denote quantities associated with \texttt{SenderReceiver}.  The factor of 2 before $\tau_{\mathrm{link}}$ accounts for Alice's photon propagating distance $L$ to Bob, followed by classical signals from Bob indicating which of her transmissions that Bob latched into memory (she learns the result $\tau_{\mathrm{link}}$ after Bob does).  The messaging is similar to \texttt{MeetInTheMiddle}, except that the locations in memory for stored entanglement are not the same for Alice and Bob.  For each arriving photon $i \in [1,N_A]$, Bob will transmit a message $\mathcal{M}_i$ to Alice that says either ``Transmission $i$ failed'' or ``Transmission $i$ is entangled with memory qubit $j$ in Bob,'' for some $j \in [1,N_B]$.   As before, we assume that both repeaters wait until the end of the round to use entanglement and that all memory qubits are reset before starting the next round.

\begin{figure}
\centering
  \includegraphics[width=8.3cm]{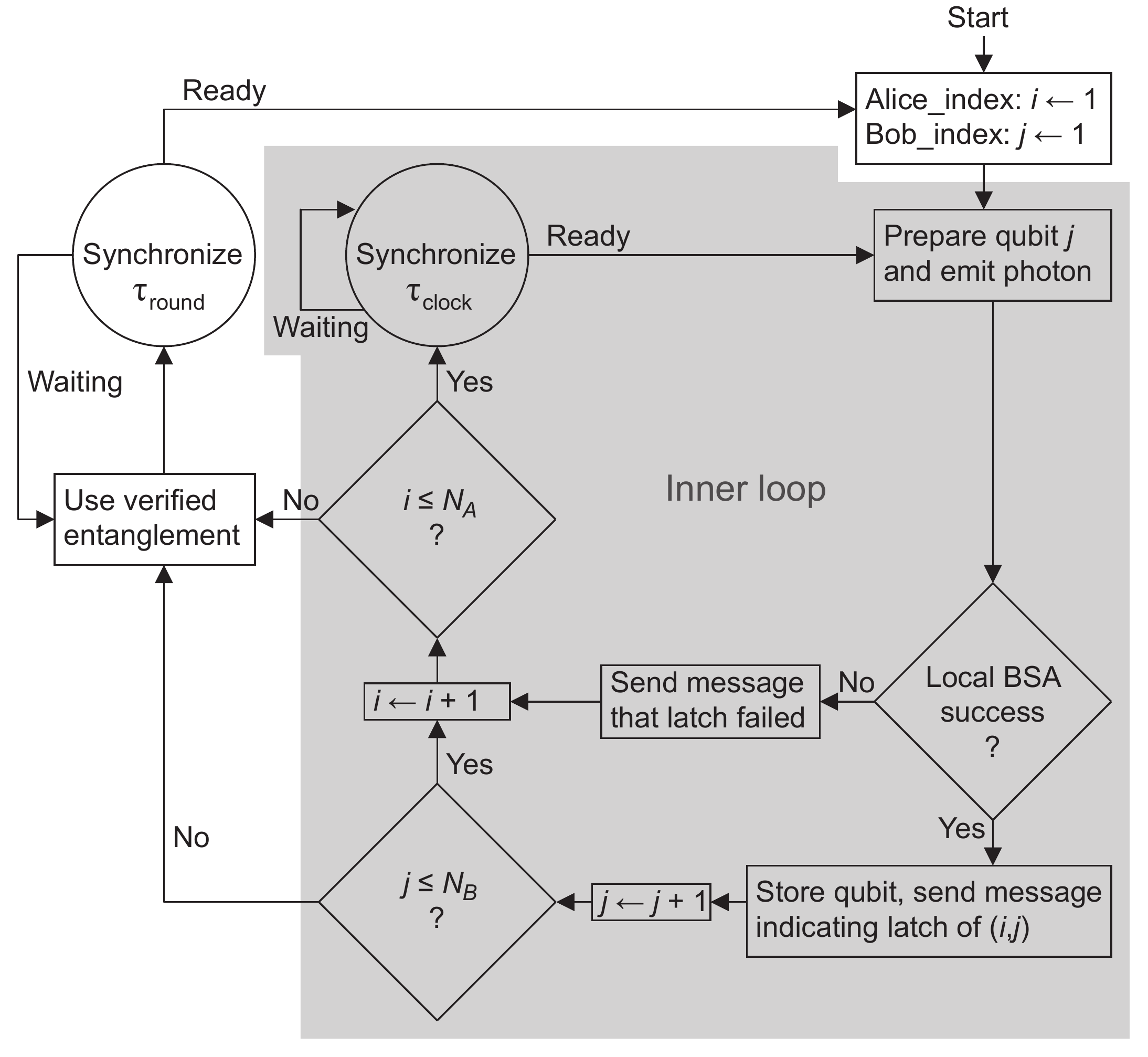}
  \caption{Control protocol for the receiver (Bob) in \texttt{SenderReceiver}.  Loop variables $i$ and $j$ follow the definitions in the text.}
  \label{fig::SR_control}
\end{figure}

The control protocol for \texttt{SenderReceiver} is different for Alice and Bob.  Alice has essentially the same control as \texttt{MeetInTheMiddle} (see Fig.~\ref{fig::MITM_control}): send a sequence of photons into the channel and wait for a response message for each after $2\tau_{\mathrm{link}}$, because she is now distance $L$ from the BSA.  However, Bob acts much faster.  His protocol, shown in Fig.~\ref{fig::SR_control}, resets a memory qubit when latching fails, allowing the next incoming photon to attempt latching into that memory.  After latching succeeds, Bob will either try to latch subsequent incoming photons into the next memory qubit or reject them if his memory is full.

The asymmetry of \texttt{SenderReceiver} forces one to consider how many memory qubits should be allocated between Alice and Bob.  Some quantum communication protocols like entanglement distillation~\cite{Gisin1995,Bennett1996_distill,Bennett1996_distill2,Kwiat2000,Zhao2003} require Bob and Alice to operate on multiple pairs of entangled qubits simultaneously, so Bob may need more than one memory qubit to access multiple entangled qubits at the same time.  In this case, \texttt{SenderReceiver} works best when the ratio of memory sizes in Bob and Alice is $N_B/N_A \approx p$, where $p$ is probability of successful latching at Bob (assumed to be the same as \texttt{MeetInTheMiddle}).  By satisfying this ratio, Alice is expected to entangle with all of Bob's memory qubits on average, while utilizing all of her memory qubits.  If the ratio is much greater or less than $p$, then one of the repeaters will not be utilizing all of its memory.

As with \texttt{MeetInTheMiddle}, we can understand the efficiency of \texttt{SenderReceiver} by the fraction of time that the protocol spends in the inner loop.  The sender and receiver have different protocols, but they spend about the same amount of time in their respective inner loops, because Bob attempts to latch every photon that Alice sends, until his memory is full.  Determining the average communication rate $R'$ for a link using \texttt{SenderReceiver} is a little more complicated since Bob may reject some of Alice's photons if his memory fills up:
\begin{align}
& R' & = & \; \frac{1}{\tau_{\mathrm{round}}'} \sum_{x=0}^{N_A} \mathrm{min}(x,N_B) \binom{N_A}{x} p^x (1-p)^{N_A-x} \label{eqn::SR_rate_exact} \\
& & < & \; \frac{p N_A}{\tau_{\mathrm{round}}'}.
\end{align}
In Eqn.~(\ref{eqn::SR_rate_exact}) the term $\mathrm{min}(x,N_B)$ accounts for the possibility that Bob rejects incoming photons after filling up his memory, and the upper bound is given by replacing the minimum function with summation variable $x$ (equivalent to assuming $N_B = N_A$).  

To make a fair comparison with \texttt{MeetInTheMiddle}, let us say that the number of memory qubits connected to a link is fixed, meaning $2N = N_A + N_B$, which is motivated by the following line of reasoning.  We assume for the moment that number of memory qubits is the limiting resource for repeater technology, so we will compare the two protocols when this quantity is fixed.  Consider a linear chain of repeaters, where each has $2N$ memory qubits and is connected to two links.  The repeaters could implement \texttt{MeetInTheMiddle}, where each repeater assigns $N$ qubits to a link.  Alternatively, the repeaters could implement \texttt{SenderReceiver}, where each repeater is a sender in one direction and a receiver in the other.  If we assume the links have identical parameters, then $N_A$ and $N_B$ will be the same for all links, and the number of qubits assigned to a link from both connected repeaters is $2N$, as illustrated with the example $N = 3$ in Fig.~\ref{fig::sender_and_receiver}.   Under these conditions, $N_A < 2N$ and
\begin{equation}
R' < \frac{p N_A}{2\tau_{\mathrm{link}} + N_A \tau_{\mathrm{clock}}} < \frac{p N}{\tau_{\mathrm{link}} + N \tau_{\mathrm{clock}}} = R.
\end{equation}
In other words, \texttt{SenderReceiver} has a lower average rate than \texttt{MeetInTheMiddle}!  However, a few considerations should be made.  First, \texttt{SenderReceiver} places the BSA inside a repeater rather than at the link midpoint, which could be quite important for practical concerns related to installing a network.  Second, one can show that $R$ and $R'$ are similar if $N_B/N_A \approx p$.  Third, in \texttt{SenderReceiver}, the receiver spends nearly the same amount of time in its inner loop as the sender despite using far fewer memory qubits.  We omit analytical derivation for optimal values of $N_A$, $N_B$, or link utilization $F'$, because such analysis is complicated and not particularly illuminating; instead, we estimate these quantities with numerical simulation in Section~\ref{sec::simulation}.  Ultimately, both \texttt{MeetInTheMiddle} and \texttt{SenderReceiver} are limited by the delays for memory lock up at the sender interface(s), so the final protocol considers what happens when all link interfaces are receivers.

\subsection{Midpoint-Source}
\label{sec::MPS}
\begin{figure*}
\centering
  \includegraphics[width=14cm]{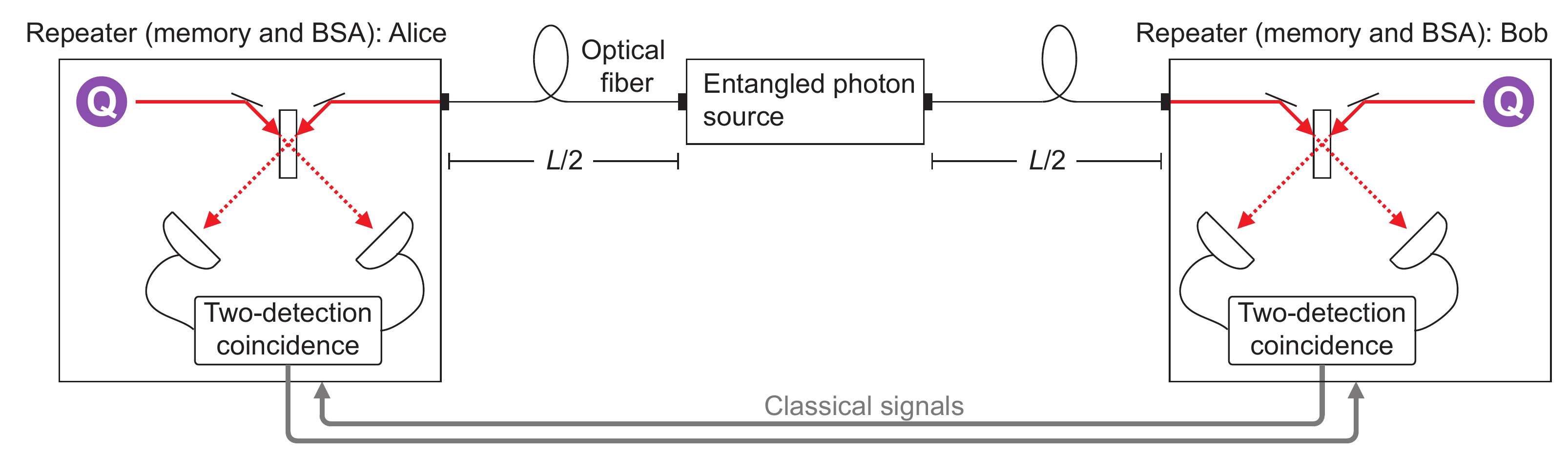}
  \caption{Hardware arrangement for \texttt{MidpointSource}.  A source of entangled photons in the middle of the channel splits a pair of entangled qubits, sending one to each repeater.  The source can generate photon pairs at a fast clock rate, so long as the repeaters are synchronized with it.  For each arriving photon, the repeater uses its protocol to determine if the photonic qubit was latched into memory using teleportation.  Each repeater node has an internal BSA, allowing rapid reset of a memory qubit if latching fails.}
  \label{fig::MPS}
\end{figure*}

The \texttt{MidpointSource} protocol, shown in Fig.~\ref{fig::MPS}, is the most complex entanglement-distribution scheme that we consider, and it uses more optical network components than the other two.  However, our analysis shows that this extra complexity is justified in many scenarios because \texttt{MidpointSource} is more robust to photon loss.  While \texttt{MeetInTheMiddle} and \texttt{SenderReceiver} have very similar performance, \texttt{MidpointSource} has a fundamentally different average rate as a function of transmission probability $p$; the former protocols have rate proportional to $p$, but \texttt{MidpointSource} has rate that scales like $\sqrt{p}$, which can be a dramatic improvement when $p \ll 1$. \texttt{MidpointSource} was introduced in Ref.~\cite{Jones2013}.

In \texttt{SenderReceiver}, the receiver node is able to make efficient use of its memory qubits by using loss heralding to know instantly whether a photon was latched into memory.  In \texttt{MidpointSource}, both repeaters connected to a link use the receiver interface to achieve efficient memory utilization.  Since both repeaters latch incoming photons into memory, entanglement is distributed using a source of entangled photons placed somewhere in the optical channel.  \texttt{MidpointSource} exploits the fact that sources of entangled photons are relatively mature technology compared to quantum memories, and experiments have demonstrated sources of high-fidelity entangled photons available at up to MHz clock rates~\cite{Honjo2007,Medic2010,Pomarico2012}.  Similar entanglement-distribution schemes have been studied that place a source of entangled photons at the link midpoint~\cite{Ekert1991,Aspelmeyer2003,Simon2007,Gisin2007,Sangouard2011}.  The \texttt{MidpointSource} protocol goes further, using fast discrimination of lost photons to achieve rapid entanglement distribution that is less sensitive to loss~\cite{Jones2013}.

To set the scene for a control protocol, consider that the latching process for a receiver in \texttt{MidpointSource} does not carry the same information as it does in \texttt{SenderReceiver}.  In \texttt{SenderReceiver}, a successful latch indicates that a photon traversed the entire channel, but a latch in \texttt{MidpointSource} only indicates that a photon from the entangled pair source has been stored in this repeater, without any indication of whether the other photon from that entangled pair was latched into the distant receiver.  Having less information per latch event might appear to put \texttt{MidpointSource} at a disadvantage, but the protocol redeems itself if the hardware can attempt latching at a very fast clock rate.  Each receiver independently tries to latch each arriving photon.  After a latch attempt succeeds, the receiver holds this qubit in memory and sends a signal to the other receiver indicating the latch.  Subsequent photons may be rejected for a short period, as described below.  If both receivers latch a photon from the same entangled-photon pair, then the corresponding memory qubits are entangled, which is confirmed by both repeaters using classical messages that require time $\tau_{\mathrm{link}}$ to propagate.  Otherwise, a stored qubit is discarded when a repeater learns that the other photon was not latched, which happens after delay of at most $\tau_{\mathrm{link}}$.

\begin{figure*}
\centering
  \includegraphics[width=14cm]{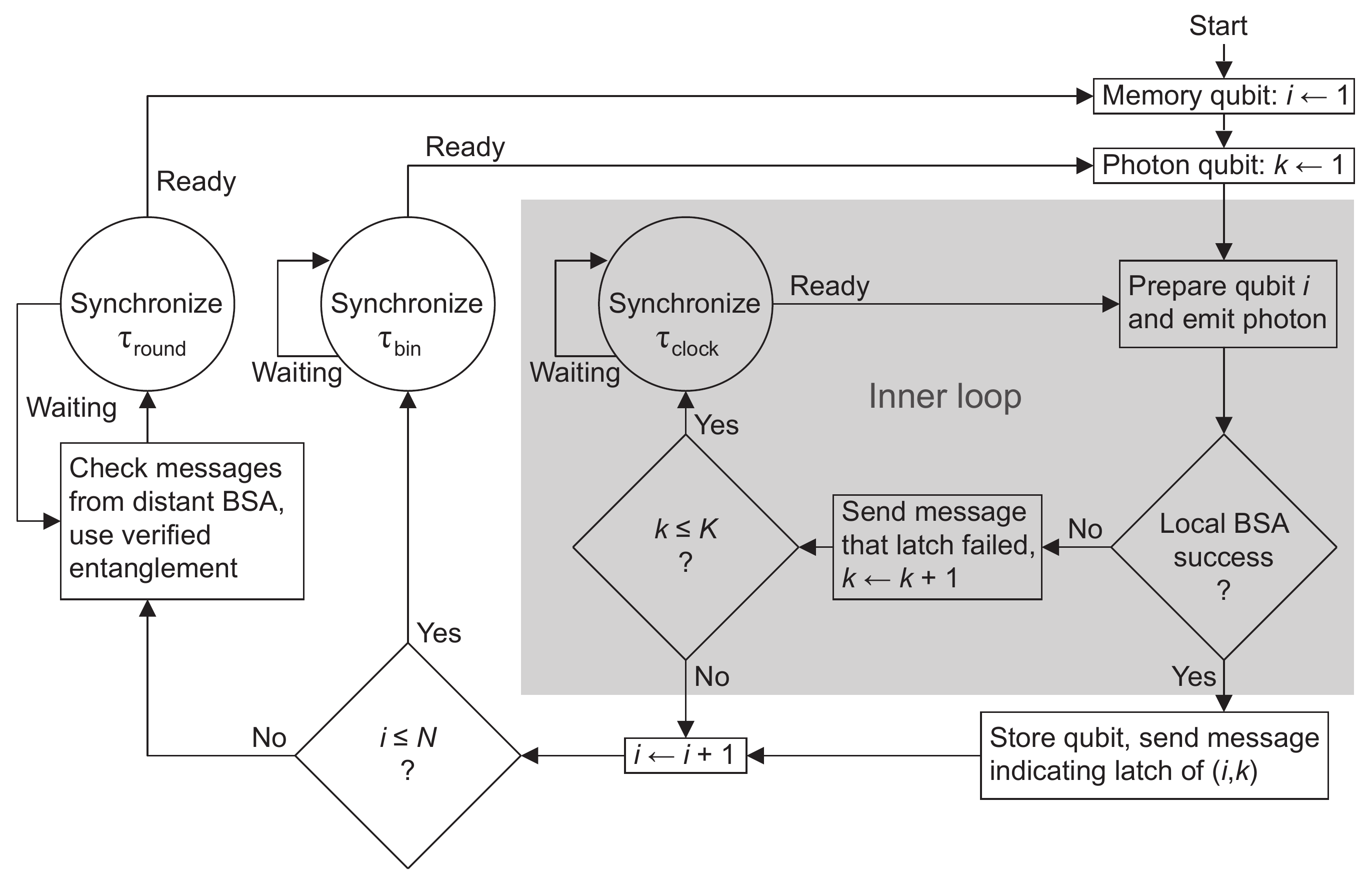}
  \caption{Control protocol for each receiver in \texttt{MidpointSource}. Loop variables $i$ and $k$ follow the definitions in the text.}
  \label{fig::MPS_control}
\end{figure*}

There are several ways to design a protocol for \texttt{MidpointSource}.  The one implemented in Fig.~\ref{fig::MPS_control} operates in discrete rounds, for direct comparison with \texttt{MeetInTheMiddle} and \texttt{SenderReceiver}.  Operating in a free-running mode with asynchronous memory reset could be more efficient, but it also requires more complex control; the case of having one qubit per receiver was solved in Ref.~\cite{Jones2013}, and specifying an asynchronous \texttt{MidpointSource} protocol for more than one memory qubit is a matter for future work.  For the protocol in Fig.~\ref{fig::MPS_control}, each repeater has $N''$ memory qubits, where double prime ($''$) denotes quantities associated with \texttt{MidpointSource}.  The round consists of a series of $N''$ communication time ``bins,'' followed by one delay of $\tau_{\mathrm{link}}$ to confirm entanglement, so
\begin{equation}
\tau_{\mathrm{round}}'' = \tau_{\mathrm{link}} + N'' \tau_{\mathrm{bin}}''.
\end{equation}
Each bin is associated with a memory qubit.  During bin $i$, the protocol attempts to latch incoming photons into memory qubit $i$.  If latching succeeds, the protocol rejects subsequent photons until the start of the next bin, when the protocol attempts to latch into the next qubit.  Within a communication bin, every entangled-photon pair from the midpoint source has a unique identifier $k \in [1, K]$, where $K$ will be determined below.  As with \texttt{SenderReceiver}, each receiver generates a message $\mathcal{M}_{i,k}$ for each bin $i$ and incoming photon $k$ in that bin.  The message says simply ``Photon $k$ was latched into memory $i$'' or ``Photon $k$ in bin $i$ was rejected.''  After attempting entanglement in all $N''$ bins, the protocol waits $\tau_{\mathrm{link}}$ to confirm entanglement.

Because the arrangement of hardware is different from the other two protocols, we need to define some new quantities for probability of latching a photon into memory.  Let $p_m''$ be the probability that an entangled-photon pair is generated at the midpoint source.  If a pair is generated, let $p_l''$ ($p_r''$) be the probability that the left (right) photon is latched into the repeater that receives it.  If there is just a single attempt at generating entanglement, the probability of success would be 
\begin{equation}
p'' = p_l'' p_m'' p_r''.  
\label{eqn::MPS_p}
\end{equation}
Note that even if $p_m'' = 1$, we expect that the probability of latching both qubits into memory would obey $p_l'' p_r'' < p$, because $p$ for \texttt{MeetInTheMiddle} and \texttt{SenderReceiver} accounts for one BSA while \texttt{MidpointSource} has two.  We assume $p_l'' = p_r''$ throughout our analysis, though the protocol can be modified for asymmetric designs.  

We further decompose loss into the fundamental pieces of a link, which will aid our comparison of protocols.  We assume that the probability of successful BSA when two photons arrive is $p_{\mathrm{BSA}}$, which is the same for all protocols.  Furthermore, $p_{\mathrm{optical}}$ is the product of probabilities for successful transmission through the memory/photon interface (with fiber coupling) and transmission through optical fiber over distance $L/2$ (half of the link distance).  We can write the total link-transmission probability for \texttt{MeetInTheMiddle} and \texttt{SenderReceiver} as
\begin{equation}
p = p_{\mathrm{bsa}} \left(p_{\mathrm{optical}}\right)^2.
\label{eqn::success_terms_MITM}
\end{equation}
For \texttt{MidpointSource}, $p_l'' = p_r'' = p_{\mathrm{bsa}} p_{\mathrm{optical}}$, so
\begin{equation}
p'' = p_m'' \left(p_{\mathrm{bsa}} p_{\mathrm{optical}}\right)^2.
\label{eqn::success_terms_MPS}
\end{equation}
The added complexity of the entangled-photon source and second BSA are manifest in the additional terms that reduce probability of successful entanglement distribution.

The latching process in \texttt{MidpointSource} is attempted at a fast repetition rate having cycle time $\tau_{\mathrm{clock}}$ that is limited only by the maximum of three time quantities: (a)~the clock period for generating entangled photons, (b)~the recovery time for BSA detectors, and (c)~the reset time for memory qubits.  This clock cycle is independent of the signaling time across the channel, so the average time to latch a qubit can be less than $\tau_{\mathrm{link}}$.  We set the number of latch attempts per bin to be
\begin{equation}
K = \frac{\tau_{\mathrm{bin}}''}{\tau_{\mathrm{clock}}} = \left\lceil \frac{3}{p_l'' p_m''} \right\rceil.
\label{eqn::clock_per_bin}
\end{equation}
The numerator 3 in Eqn.~(\ref{eqn::clock_per_bin}) is selected somewhat arbitrarily to make the probability of latching a photon in a time bin near unity: 
\begin{equation}
p_{\mathrm{latch}} = 1 - (1-p_l'' p_m'')^K > 0.95.
\label{eqn::latch_per_bin}
\end{equation}
We will use Eqn.~(\ref{eqn::latch_per_bin}) repeatedly to place bounds on the average-case performance of \texttt{MidpointSource}.

The key advantage of \texttt{MidpointSource} is that a latched qubit has already overcome at least half of the loss for one side of the link.  In particular, the probability that both repeaters latch a photon in time bin $i$ is greater than $0.90$ with the above parameters.  Memory qubits at index $i$ for two receivers are entangled \emph{only} if they both latched photons from entangled pair $k$.  For the left repeater, the probability that a latched memory qubit is entangled to its partner on the right side is a little complicated, since latching is only attempted for photon $k$ if latching failed for photons $1$ to $k-1$.  We can express the probability that entanglement is established in time bin $i$ as the sum:
\begin{equation}
p_{\mathrm{ent}}'' = \sum_{k=1}^K p'' \left[1 - p_m''(p_l'' + p_r'') + p''\right]^{k-1}.
\end{equation}
Using Eqn.~(\ref{eqn::MPS_p}), this can be expressed as
\begin{align}
p_{\mathrm{ent}}'' & = & & p'' \sum_{k=1}^K \left(1 - p'' (\frac{1}{p_l''} + \frac{1}{p_r''} - 1)\right)^{k-1} \nonumber \\
& = & & \frac{p_l''}{2 - p_l''} \left[1 - \left(1 - p'' (\frac{2}{p_l''} - 1)\right)^K \right],
\end{align}
where we assume as before that $p_l'' = p_r''$.  Using Eqn.~(\ref{eqn::latch_per_bin}) and the trivial fact that $1/p_l'' > 1$, we can derive bounds on the entanglement probability:
\begin{equation}
0.95 \frac{p_l''}{2} < p_{\mathrm{ent}}'' < \frac{p_l''}{2 - p_l''}. 
\end{equation}
If $p_l'' < 0.1$, then we can say that $p_{\mathrm{ent}}'' \approx p_l''/2$ with at most 5\% relative error.  

The average rate of entanglement is simply the expected number of entanglement events per round:
\begin{equation}
R'' = \frac{N'' p_{\mathrm{ent}}''}{\tau_{\mathrm{round}}''}.
\end{equation}
We can compare this rate to \texttt{MeetInTheMiddle} (which bounds \texttt{SenderReceiver}) using the decomposition into success terms in Eqns.~(\ref{eqn::success_terms_MITM}) and~(\ref{eqn::success_terms_MPS}).  We set $N'' = N$ and assume that $N \tau_{\mathrm{bin}}'' \ll \tau_{\mathrm{link}}$ (``fast-clock'' assumption, discussed below), so
\begin{equation}
R'' \approx \frac{N p_{\mathrm{bsa}} p_{\mathrm{optical}}}{2\tau_{\mathrm{link}}}. 
\end{equation}
Under the same fast clock assumption,
\begin{equation}
R \approx \frac{N p_{\mathrm{bsa}} \left(p_{\mathrm{optical}}\right)^2}{\tau_{\mathrm{link}}}.
\label{eqn::MS_rate}
\end{equation}
The ratio of the two is
\begin{equation}
\frac{R''}{R} \approx \frac{1}{2 p_{\mathrm{optical}}}.
\end{equation}
The probability $p_{\mathrm{optical}}$ is associated with memory/photon interface losses and fiber attenuation, and is common to all protocols.  In particular, we can say that $R''$ is scaling like $p_{\mathrm{optical}}$ while $R$ and $R'$ are scaling as $\left(p_{\mathrm{optical}}\right)^2$, explaining the $\sim \sqrt{p}$ separation mentioned at the beginning of this section.  Another interpretation is that the \texttt{MidpointSource} protocol is less sensitive to the underlying signal losses represented by $p_{\mathrm{optical}}$, making \texttt{MidpointSource} well-suited to early prototypes of quantum repeaters.

The rate in Eqn.~(\ref{eqn::MS_rate}) has a remarkable feature.  The average communication rate is independent of the probability that the entangled source generates an entangled pair, even if $p_m'' \ll 1$.  As a result, \texttt{MidpointSource} works very well, even if a probabilistic source of entangled photons is used, such as spontaneous parametric downconversion.  The explanation for this effect has two components.  First, we choose a number of latching attempts per round ($K$) according to Eqn.~(\ref{eqn::clock_per_bin}), which scales inversely with $p_m''$.  Under the fast-clock assumption, the time for latching attempts is insignificant compared to total round time, indicating how critical a fast clock is for ``absorbing'' the impact of low signal transmission probability.  Second, we presume that the entangled-pair source does not emit single photons; the fact that a receiver latches post-selects events where a photon was sent to the other receiver.  We set $K$ so that a receiver latches with near certainty, and when that receiver latches, the probability that the other receiver latched a photon from the same pair is proportional to $p_{\mathrm{optical}}$.  In addition to having less sensitivity to loss in optical components, \texttt{MidpointSource} has almost no dependence on $p_m''$, showing how robust the protocol is to signal loss.  However, we emphasize that this robustness depends on a fast clock cycle at the receiver.

Another way to see how \texttt{MidpointSource} achieves high performance is the fraction of time spent in its inner loop.  In a time bin, this fraction is
\begin{equation}
F_{\mathrm{bin}}'' = \frac{1}{K}\sum_{k=1}^K k p_l'' p_m'' (1 - p_l'' p_m'')^k,
\end{equation}
and we use Eqn.~(\ref{eqn::latch_per_bin}) to establish bounds $\frac{0.95}{3} < F_{\mathrm{bin}}'' < \frac{1}{3}$.  The link utilization for one round of \texttt{MidpointSource} is
\begin{equation}
F'' = \frac{N'' F_{\mathrm{bin}}'' \tau_{\mathrm{bin}}''}{\tau_{\mathrm{round}}''}.
\end{equation}
However, the connection between utilization and rate is not as simple as before: $R'' \neq F'' p''/ \tau_{\mathrm{clock}}''$, due to the following difference in available information.  When the receiver in \texttt{SenderReceiver} exits its inner loop, it knows that any latch events correspond to established entanglement, whereas when \texttt{MidpointSource} exits its inner loop, there is only a probability about $p_l'' /2$ that both receivers latched the same photon.  \texttt{MidpointSource} compensates by having a rate of photon arrival from the entangled-photon source that is independent of the memory size in either repeater, which is not true for \texttt{MeetInTheMiddle} or \texttt{SenderReceiver}.

We should consider for a moment the fast-clock assumption made above.  This simplified our analysis, but satisfying this condition is also necessary for \texttt{MidpointSource} to yield higher communication rate.  One can show that the performance of \texttt{MidpointSource} degrades to being worse than the other two protocols in the other limit, $N \tau_{\mathrm{bin}}'' > \tau_{\mathrm{link}}$.  Under the fast clock assumption, memory qubits are locked up for about $\tau_{\mathrm{link}}$ in \texttt{MeetInTheMiddle} and \texttt{MidpointSource}.  When this assumption does not hold, the protocols as specified will tend to produce entanglement at a rate independent of $N$, because memory qubits are locked up for round times that scale with $N$.  Worse yet, \texttt{MidpointSource} has lower success probability per entanglement attempt due the additional entangled-photon source and BSA, and a fast clock is essential to offset this complexity.  In summary, all protocols benefit from fast clock cycle $\tau_{\mathrm{clock}} \ll \tau_{\mathrm{link}}/N$, and \texttt{MidpointSource} shows the greatest benefit when the clock period satisfies $\tau_{\mathrm{clock}} \ll \tau_{\mathrm{link}}/(NK)$.

Since \texttt{MidpointSource} depends critically on the fast clock condition, we present a substitute for the entangled-pair source that can realize $p_{\mathrm{mid}}'' = 0.5$, thereby reducing $K$ and easing the requirements on $\tau_{\mathrm{clock}}$.  Suppose two triggered, deterministic single-photon sources emit photons that are indistinguishable except for polarization, where one emits horizontal and the other vertical.  These photons interfere on a beamsplitter, and the photons exiting the two output ports are collected.  This approach was presented in Ref.~\cite{Jones2013} as an alternative to an entangled-photon source for \texttt{MidpointSource}.  Label the left/right input modes of the beamsplitter as ``a''/``b'', and label the left/right output modes as ``c''/``d''.  When the single photons interfere at the beamsplitter, the state is transformed as
\begin{equation}
\ket{H}_a \ket{V}_b \rightarrow \frac{1}{\sqrt{2}}\ket{\Psi^-} - \frac{1}{2} \ket{H}_c \ket{V}_c + \frac{1}{2} \ket{H}_d \ket{V}_d,
\end{equation}
where
\begin{equation}
\ket{\Psi^-} = \frac{1}{\sqrt{2}}\left(\ket{H}_c \ket{V}_d - \ket{V}_c \ket{H}_d\right)
\end{equation}
is a maximally entangled state of the two photons.  Note that the entangled state can trigger BSA success at both receivers, whereas the states $\ket{H}_c \ket{V}_c$ and $\ket{H}_d \ket{V}_d$ cannot.  As a result, the interference of two single-photon sources is a ``post-selected'' source of entanglement~\cite{Fattal2004} when used in the \texttt{MidpointSource} scheme, where the post-selection occurs when both BSAs indicate successful latching.  The entangled-photon state occurs in half of the attempts. 

\texttt{MidpointSource} is more complex than both \texttt{MeetInTheMiddle} and \texttt{SenderReceiver} in two ways.  First, the link requires more optical hardware.  There are two BSAs, meaning more single-photon detectors, and there is a source of entangled photons.  However, both the preceding analysis and numerical simulation in Section~\ref{sec::simulation} show that the additional optical components allow \texttt{MidpointSource} to make more efficient use of memory qubits.  Consequently, a network employing \texttt{MeetInTheMiddle} or \texttt{SenderReceiver} would require more memory qubits in one or both repeaters to match the performance of \texttt{MidpointSource}.  Given that memory qubits are currently a less mature technology than either entangled-photon sources or single-photon detectors, \texttt{MidpointSource} might be the best protocol for early repeater networks because it requires fewer memory qubits at the expense of more optical components.  This trade-off in resources motivates our numerical simulations, and we make that trade-off more quantitative in Section~\ref{sec::simulation}.  The second way that \texttt{MidpointSource} is more complex is that its control protocol has much more information to process.  Nevertheless, the demands that the protocol in Fig.~\ref{fig::MPS_control} places on both local processors and network transmission seem modest for classical information technology, and we argue that this additional complexity is justified by higher rate of entanglement distribution for the same number of memory qubits.

\subsection{Further Development of Protocols}
The protocols developed in this manuscript were designed to demonstrate key features of the three ways of linking two repeaters (sender-sender, sender-receiver, receiver-receiver).  However, they are not optimal, and further performance improvements are possible.  For example, waiting $\tau_{\mathrm{link}}$ for classical messages to propagate after all photon transmissions is not necessary in most cases.  Similarly, \texttt{MidpointSource} can operate in a ``free running'' mode that does not associate memory qubits with time bins.  Additional care must be taken to ensure that memory qubits do not get stuck in a pathological pattern of locking up asynchronously, which would prevent entanglement distribution; the case for $N = 1$ was solved in Ref.~\cite{Jones2013}.  Our protocols were designed to be simple while capturing the essential way in which performance is limited by the quantum hardware.  When the fast-clock assumption holds, these simple protocols deliver nearly optimal performance since any entanglement must be confirmed with classical signals requiring delay $\tau_{\mathrm{link}}$.  

There are interesting avenues to explore in developing better protocols.  For example, the location of a BSA in the link can determine the frequency of the interfering photons, thereby affecting the performance of single-photon detectors.  Another approach is to consider asynchronous designs that do not have a fixed round time, which could perform much better when the fast-clock approximation does not hold.  Furthermore, a round time that is longer than memory lifetimes would be unacceptable, which could be relevant if the number of memory qubits is very large.  Finally, sending multiple signals in parallel (such as with frequency-division multiplexing) can increase communication rate, though it might require a more sophisticated protocol.

The protocols considered here are not an exhaustive list, and one could search for new hardware arrangements and control schemes not yet discovered.  One way to find a new protocol would be to combine elements from the three protocols above, then eliminate any unnecessary components.  Indeed, \texttt{MidpointSource} was derived from \texttt{SenderReceiver} in the following way.  Take a repeater chain consisting of \texttt{SenderReceiver} protocol on each link, but alternate the direction (much like the ``butterfly arrangement'' in Ref.~\cite{Munro2010}).  Both link interfaces for every odd-numbered repeater are type sender in each direction, and the even-numbered repeaters have receiver interfaces.  Now consider a single sender-type node.  It simultaneously sends photons in both directions to different receivers.  Instead of holding the memory qubits while waiting for latching results from the receivers, the sender node could perform an immediate entanglement swap of these memory qubits and send out the classical result of the Bell measurement.  If the swap is executed immediately, the sender node only requires two memory qubits, since they are reset before the next photon transmission.  The sender node is simply acting as a source of entangled photons.  By replacing the sender node with a more conventional entangled-photon source that does not require quantum memory at all, you have the \texttt{MidpointSource} hardware arrangement.  It may be possible to derive new communication protocols using similar techniques of mixing and replacing fundamental repeater elements.

\section{Simulation of Network Performance}
\label{sec::simulation}
We develop numerical simulations of a quantum repeater network based on the protocols in Section~\ref{sec::link_design}, to compare their performance using more complicated models that do not admit simple analytical results.  We perform two types of comparisons.  First, we compare the protocols in a repeater network consisting of ten links using a common set of parameters that represent a mature platform for repeater technology, including long-lived quantum memory and low-error gates for purification.  Moreover, we choose parameters that satisfy the fast-clock condition to highlight the differences between protocols.  The repeaters store successfully transmitted entanglement in memory and perform purification, as explained below.  The results are straightforward: \texttt{MidpointSource} is the best protocol when the fast-clock assumption is valid.  If this assumption does not hold, the simpler \texttt{MeetInTheMiddle} may perform better.  

In the second simulation, we evaluate the performance of the protocols for near-term experiments using current state-of-the-art device parameters.  Two nodes establish entanglement across a single link and perform immediate measurement, and we compare the rate of entanglement generation for \texttt{MeetInTheMiddle} and \texttt{MidpointSource}.  The three repeater technologies we consider are trapped ions, diamond NV centers, and quantum dots; these devices can store qubits in memory, apply operations, and interface memory with single photons.  For this set of simulations, the fast-clock assumption does not necessarily hold, and we can assess how well these protocols might perform in practice.

\subsection{Simulations to Compare Protocols}
The model network in our simulations is a linear chain of ten links (eleven repeater nodes), which distribute entanglement across each link with initial fidelity of 0.95.  We purify entangled pairs using the decoding circuit of the $[[7,1,3]]$ Steane code~\cite{Nielsen2000,Knill2005}, where we assume for simplicity that local gates and memory in the repeater are error-free.  In a more realistic setting, errors can be suppressed using quantum error correction~\cite{Nielsen2000,Knill2005}, but a detailed implementation here is a matter for future work.  The error on each input Bell state to the purification procedure is i.i.d. $\epsilon_{\mathrm{in}}$, so the error of a successfully purified state is bounded by
\begin{equation}
\epsilon_{\mathrm{out}} \le 7 \epsilon_{\mathrm{in}}^3 (1-\epsilon_{\mathrm{in}})^4 + \epsilon_{\mathrm{in}}^7,
\end{equation}
and the probability of success is bounded by
\begin{equation}
p_{\mathrm{success}} > (1-\epsilon_{\mathrm{in}})^7.
\end{equation}
For $\epsilon_{\mathrm{in}} = 0.05$, we have $\epsilon_{\mathrm{out}} < 10^{-3}$ and $p_{\mathrm{success}} > 0.698$.  After entanglement purification across each link, the network establishes end-to-end entangled pairs using entanglement swapping, with total error over ten links bounded by $\epsilon_{\mathrm{total}} < 10 \epsilon_{\mathrm{out}} < 10^{-2}$.  The end-to-end entangled pairs with fidelity greater than 0.99 can be used for QKD.  The communication rate of the network is reported as the number of these end-to-end entangled qubit pairs (ebits) created per second.  The communication rate is plotted as a function of inter-repeater link distance.  Each plotted point is the mean of 1000 samples taken, and error bars show the 90\% confidence interval for the sampled distribution.

\begin{figure}
\centering
  \includegraphics[width=8.3cm]{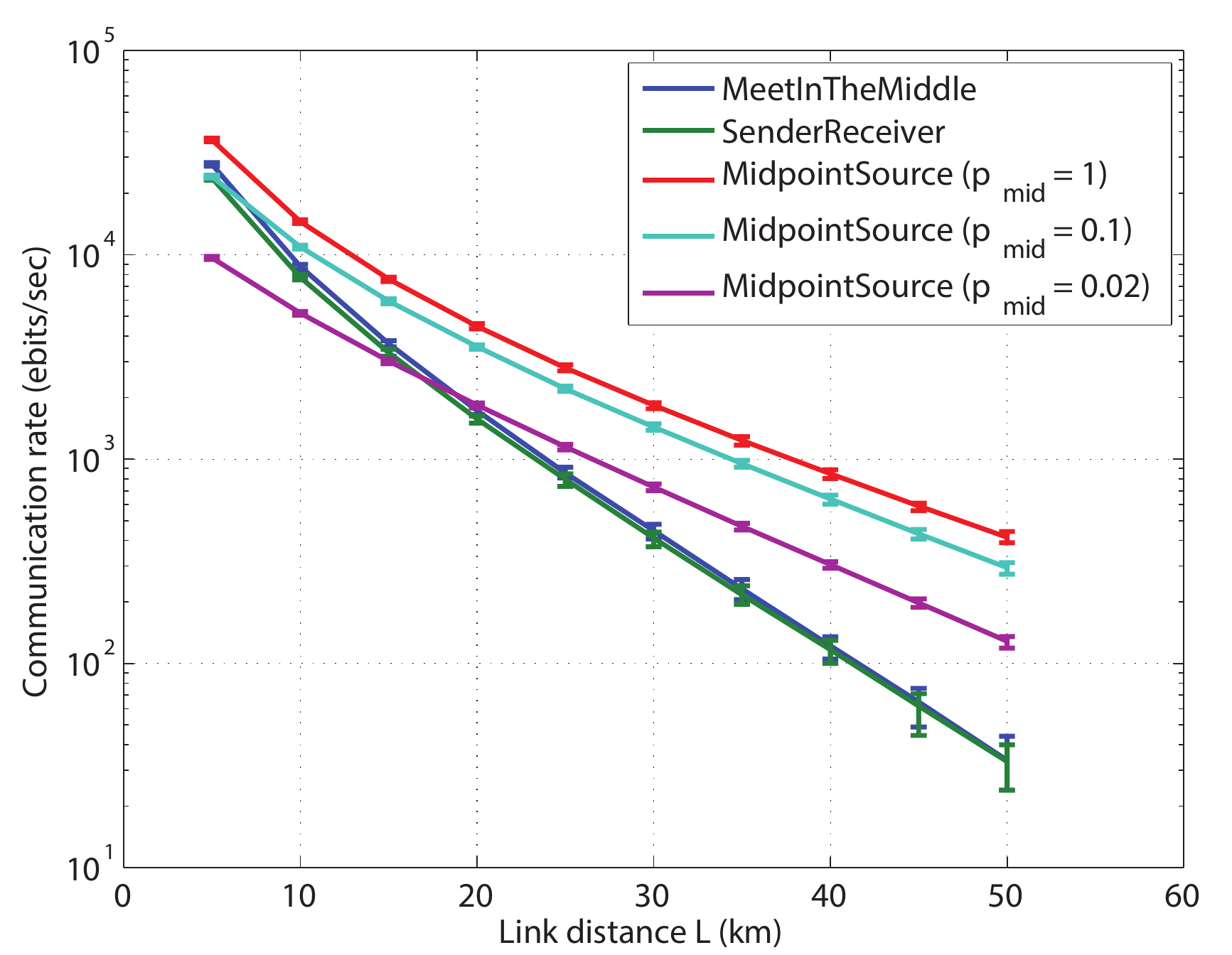}
  \caption{Communication rate for the link protocols with optimistic hardware parameters, meaning high transmission probability, as a function of link distance $L$.  There are ten links, so network length is $10L$.  The parameters are: $N = 100$, $p_{\mathrm{BSA}} = 0.5$, $p_{\mathrm{optical}} = 0.5 \exp(-L/2L_{\mathrm{att}})$, $\tau_{\mathrm{clock}} = 1$~ns.}
  \label{fig::sim_optimistic}
\end{figure}

The first simulation in Fig.~\ref{fig::sim_optimistic} uses an ``optimistic'' set of parameters, where the linear-optics BSA has maximum success probability 50\% (meaning perfect single-photon detectors), each memory-photon interface has transmission probability 50\%, and optical fiber has standard attenuation length of $L_{\mathrm{att}} = 22$~km (0.2~dB/km).  The number of memory qubits is given by $N = 100$, and the clock time is 1~ns, which was chosen to illustrate the impact of the fast-clock assumption.  Each repeater has three of its memory qubits reserved for storing purified entangled states waiting to be swapped, while the rest participate in attempting to establish entanglement.  Since the purification protocol requires seven qubits, the number of receiver qubits in \texttt{SenderReceiver} is chosen to be $6 + \lceil 2 N p/(p+1)\rceil$, where $p = p_{\mathrm{BSA}}(p_{\mathrm{optical}})^2 $ (see Section~\ref{sec::SR}).  The total simulation time is $10^3 \tau_{\mathrm{link}}$, which depends only on inter-repeater distance $L$.  We simulate three values for the entangled-photon generation probability in \texttt{MidpointSource} ($p_{\mathrm{mid}}'' = 1$, 0.1, and 0.02). The first two values satisfy the fast-clock assumption of Section~\ref{sec::link_design} since $\tau_{\mathrm{link}}$ ranges from 25 to 250~$\mu$s ($L = 5$ to 50~km), while $NK\tau_{\mathrm{clock}}$ for \texttt{MidpointSource} ranges from 700~ns to 2~$\mu$s for $p_{\mathrm{mid}}'' = 1$, and this time is 10 and 50 times larger for $p_{\mathrm{mid}}'' = 0.1$ and 0.02, respectively.  For $p_{\mathrm{mid}}'' = 0.02$, the fast-clock assumption is violated, and the performance of \texttt{MidpointSource} is degraded.  Nevertheless, the robustness of \texttt{MidpointSource} to signal loss allows even the $p_{\mathrm{mid}}'' = 0.02$ instance to outperform the other two protocols for $L \ge 20$~km.

The simulation results in Fig.~\ref{fig::sim_optimistic} show several features that are consistent with the analysis in Section~\ref{sec::link_design}.  First, \texttt{MeetInTheMiddle} and \texttt{SenderReceiver} have very similar performance, with the former being slightly better.  Second, \texttt{MidpointSource} has a communication rate that decreases with smaller slope (less dependence on transmission probability) than \texttt{MeetInTheMiddle}, because the high clock rate of \texttt{MidpointSource} enables it to be less sensitive to photon loss.  Whether \texttt{MidpointSource} outperforms \texttt{MeetInTheMiddle} depends on the link distance and the probability that an entangled-pair source generates a photon pair, which determines in part whether the fast-clock condition is satisfied.

\begin{figure}
\centering
  \includegraphics[width=8.3cm]{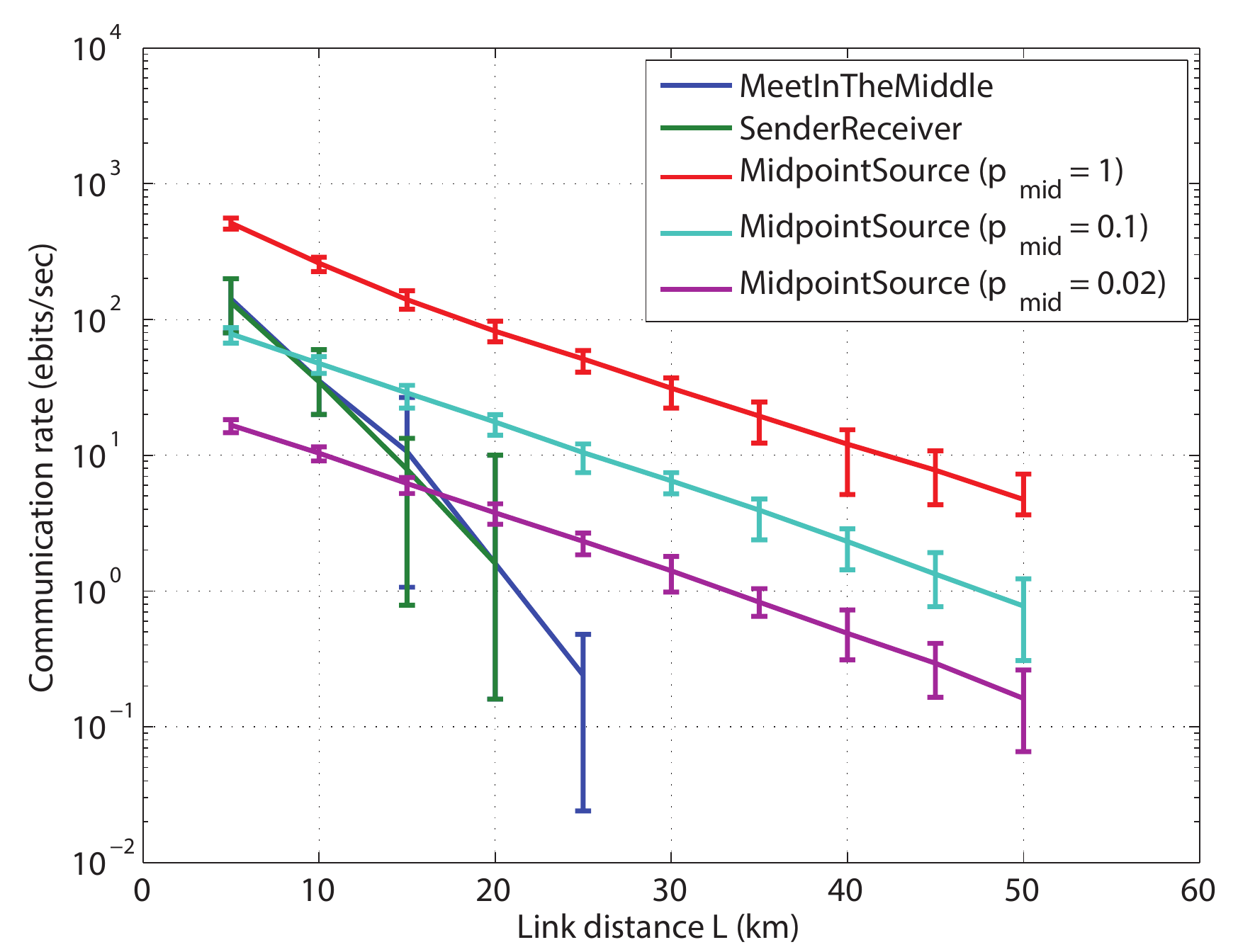}
  \caption{Communication rate for the link protocols with pessimistic hardware parameters, meaning low transmission probability, as a function of link distance $L$.  There are ten links, so network length is $10L$.  The parameters are: $N = 100$, $p_{\mathrm{BSA}} = 0.1$, $p_{\mathrm{optical}} = 0.1 \exp(-L/2L_{\mathrm{att}})$, $\tau_{\mathrm{clock}} = 1$~ns.  The downward curve in rate for \texttt{MeetInTheMiddle} and \texttt{SenderReceiver} for $L \ge 20$~km is a consequence of finite simulation time for Markov-chain Monte Carlo.  End-to-end entangled pairs require entanglement distribution across all links, and these protocols generate link-level pairs so slowly that the number of entangled pairs in the repeater memories has not reached a steady-state distribution for the finite time of the simulation, which is $10^3 \tau_{\mathrm{link}}$ (25 to 250~$\mu$s).}
  \label{fig::sim_pessimistic}
\end{figure}

Another simulation is performed for a ``pessimistic'' set of parameters, where BSA and memory/photon interface each have transmission probability 0.10, with results plotted in Fig.~\ref{fig::sim_pessimistic}.  Since photon loss is more severe, all protocols have lower communication rate than the parameter set in Fig.~\ref{fig::sim_optimistic}.  Notably, \texttt{MidpointSource} does not decrease in performance as much as the other protocols, and the gap in performance between \texttt{MidpointSource} and \texttt{MeetInTheMiddle} is generally larger than the results in Fig.~\ref{fig::sim_optimistic}.  

Unlike the ``optimistic'' set of parameters, the communication rates for \texttt{MidpointSource} protocols in Fig.~\ref{fig::sim_pessimistic} do not properly satisfy the fast-clock condition due to the lower transmission probabilities.  The result is that the communication-rate curves for different values of $p_{\mathrm{mid}}''$ are further apart in Fig.~\ref{fig::sim_pessimistic} than Fig.~\ref{fig::sim_optimistic}, because the way $K$ is chosen for these protocols (see Eqn.~\ref{eqn::clock_per_bin}) leads to the latching process accounting for most of the round duration.  Nevertheless, the fast clocking of \texttt{MidpointSource} still provides some ability to overcome signal loss, and these protocols outperform \texttt{MeetInTheMiddle} in most cases.  Notice that \texttt{MidpointSource} still has effective entanglement distribution at inter-repeater link distances up to 50~km, while both \texttt{MeetInTheMiddle} and \texttt{SenderReceiver} drop to zero end-to-end ebits around $L = 25$~km due to finite simulation time (see caption of Fig.~\ref{fig::sim_pessimistic}).

\subsection{Hardware-Specific Simulations}
The preceding simulations indicate that \texttt{MidpointSource} is the best simulation if the fast-clock condition is satisfied.  However, the clock cycle of 1~ns is very fast, and existing proposals for quantum repeater hardware do not yet operate at this speed.  We now seek to determine what the best protocol would be for ``realistic'' hardware parameters.  We consider trapped ions~\cite{Moehring2007,Kim2011,Sterk2012,Maunz2012}, diamond nitrogen-vacancy (NV) centers~\cite{Faraon2011,VanDerSar2011,Dolde2013,Bernien2013}, and quantum dots~\cite{Davanco2011,Gao2012,DeGreve2012,Schaibley2013,Arcari2014}.  Note that self-assembled quantum dots are more challenging than ions to integrate into coupled arrays, but such integration is needed to effectively integrate the communication protocols.  See Ref.~\cite{Kim2015} for a hardware proposal combining demonstrated quantum dot spin-photon methods with demonstrated methods for constructing multi-qubit arrays.  For each hardware platform, we take the optimistic approach of finding the best parameters from recent experimental results and presuming that these may be realized in one system.  The relevant parameters for clocking and memory/photon interface transmission probability are listed in Table~\ref{tbl::memory_parameters}.

\begin{table}
\centering
\caption{Timing Parameters for Memory-Photon Interfaces}
	\begin{tabular}{m{3cm} m{1.5cm} m{1.9cm} m{1.9cm}}
	\label{tbl::memory_parameters}
	\textbf{Memory type} & \textbf{Cycle time} & \textbf{Emission fraction} & \textbf{Collection efficiency}  \\ \hline
	\raggedright{Trapped ion ($^{171}$Yb$^+$)}			& 1 $\mu$s 		& 1.00  	& 0.05 \\  
	Diamond NV 								& 100 ns 		& 0.05 		& 0.50 \\ 
	\raggedright{Quantum dot (InGaAs)}				& 10 ns 		& 1.00 		& 0.50 \\ \hline
\end{tabular}
\end{table}

In addition to the quantum repeater hardware, all link protocols considered here require a BSA, and \texttt{MidpointSource} requires two BSAs and an entangled-pair source.  We assume that the BSA uses superconducting nanowire detectors (SNSPDs)~\cite{Hadfield2009,Marsili2013}.  In our model, these detectors have quantum efficiency of 0.80 and very low dark count rates, so the linear-optics BSA has success probability $p_{\mathrm{BSA}} = 0.24$.  The entangled-photon source could be one of several potential designs.  A quantum dot could potentially emit entangled photons with success probability approaching unity ($p_{\mathrm{mid}}'' = 1$)~\cite{Dousse2010}.  Two deterministic single-photon sources mixing on a beamsplitter (see Section~\ref{sec::MPS}) would produce entangled photons with $p_{\mathrm{mid}}'' = 0.5$, post-selected by the two BSAs.  Finally, entangled-photon pairs can be generated using four-photon scattering in optical fiber with $p_{\mathrm{mid}}'' = 0.02$~\cite{Li2006} or other optical nonlinearities for parametric downconversion~\cite{Kwiat1999,Honjo2007,Medic2010,Pomarico2012,Christensen2013}.  We assume that these components operate at photon frequency near 1550~nm for low-loss transmission in optical fiber.  Importantly, none of the quantum memory technologies in Table~\ref{tbl::memory_parameters} emit at this frequency, so some form of photonic frequency conversion~\cite{Rakher2010,Zaske2012,Ates2012,Kuo2013} is necessary for the interface to optical fiber.  Although important, analyzing frequency conversion is outside the scope of our work, and for this investigation we assume that signal losses associated with frequency conversion are included in the ``collection efficiency'' for each technology.

\begin{figure*}
\centering
  \includegraphics[width=17cm]{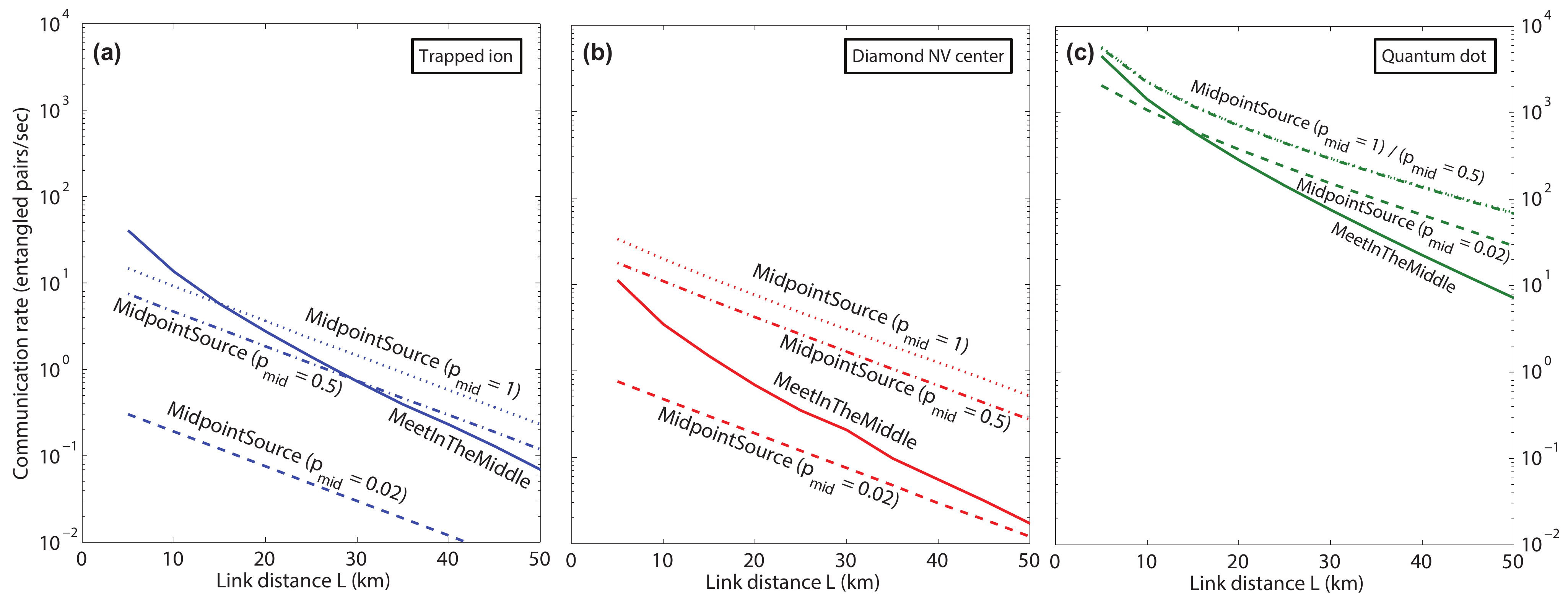}
  \caption{Communication rate for the link protocols as a function of link distance $L$ using experimentally motivated hardware parameters for (a)~trapped ions, (b)~diamond NV centers, and (c)~quantum dots.  The parameters used in this simulation correspond to Table~\ref{tbl::memory_parameters}.  The simulated network consists of just two nodes sharing one link.  The common parameters are: $N = 3$ and $p_{\mathrm{BSA}} = 0.24$.  Hardware-specific parameters are (a)~ion: $p_{\mathrm{optical}} = 0.05 \exp(-L/2L_{\mathrm{att}})$, $\tau_{\mathrm{clock}} = 1$~$\mu$s; (b)~NV: $p_{\mathrm{optical}} = 0.025 \exp(-L/2L_{\mathrm{att}})$, $\tau_{\mathrm{clock}} = 100$~ns; (c)~QD: $p_{\mathrm{optical}} = 0.5 \exp(-L/2L_{\mathrm{att}})$, $\tau_{\mathrm{clock}} = 10$~ns.  Note that in (c), the \texttt{MidpointSource} protocols with $p_{\mathrm{mid}}'' = 1$ and $p_{\mathrm{mid}}'' = 0.5$ are indistinguishable, indicating that the fast-clock condition is satisfied.}
  \label{fig::hardware_sim}
\end{figure*}

The results of the ``hardware-specific'' simulations are shown in Fig.~\ref{fig::hardware_sim}, where each of the panels corresponds to a particular repeater technology with parameters given in Table~\ref{tbl::memory_parameters}.  The plots compare \texttt{MeetInTheMiddle} to \texttt{MidpointSource} with a selection of entangled-photon sources with different values for entanglement-generation probability $p_{\mathrm{mid}}''$.  In general, the signal transmission probabilities are lower and clock rates are higher than the preceding simulations, indicating that further improvements in repeater technology are needed to realize high performance networks.  The simulations in Fig.~\ref{fig::hardware_sim} are for a single link between just two nodes.  The number of memory qubits is $N = 3$, and no purification is performed.  \texttt{SenderReceiver} has rate lower than \texttt{MeetInTheMiddle}, so it is not simulated.  The simulation runs for $10^4 \tau_{\mathrm{link}}$ to better resolve low communication rates.

The device parameters in Table~\ref{tbl::memory_parameters} were chosen to represent possible near-term experiments showing entanglement distribution to validate repeater technology.  We do not simulate memory errors, to separate comparison of the optical protocols from considerations of memory lifetime and the implementation of error correction.  We note that independence of communication rate and memory lifetime can be realized in an entanglement-tomography experiment by immediately measuring a memory qubit after either emitting a photon (\texttt{MeetInTheMiddle}) or latching (\texttt{MidpointSource}), then post-selecting cases where the BSA measurements indicate entanglement success (sometimes known as a ``delayed choice'' experiment~\cite{Scully1982,DeGreve2012,DeGreve2013}).

For many combinations of device parameters, \texttt{MidpointSource} delivers the highest communication rate, but not always.  \texttt{MeetInTheMiddle} may perform better when the fast-clock condition does not hold, and this condition becomes more difficult to satisfy with slower clock cycle or low values of $p_{\mathrm{mid}}''$.  For example, trapped ions have the slowest $\tau_{\mathrm{clock}}$, and \texttt{MidpointSource} only outperforms \texttt{MeetInTheMiddle} in Fig.~\ref{fig::hardware_sim}(a) for high values of $p_{\mathrm{mid}}''$ and link distance greater than about 20 to 30~km.  Diamond NV centers have higher collection efficiency, but they only emit about 5\% of the time into the zero-phonon line (although there are significant results showing Purcell enhancement~\cite{VanDerSar2011}, the cavity would need to be degenerate for the two photonic qubit states), so $p_{\mathrm{optical}}$ for NV centers is similar to that of ions.  However, NV centers do operate at a $10\times$ faster clock rate in our model, so \texttt{MidpointSource} shows more substantial advantage in Fig.~\ref{fig::hardware_sim}(b) when $p_{\mathrm{mid}}''$ is high.  Finally, quantum dots have the fastest clock cycle and reasonably high transmission into optical fiber; indeed, two of the \texttt{MidpointSource} curves in Fig.~\ref{fig::hardware_sim}(c) are indistinguishable, indicating that the fast-clock condition is satisfied for those parameters.  Further evidence of the benefit \texttt{MidpointSource} derives from a fast clock cycle is that the curve for $p_{\mathrm{mid}}'' = 0.02$ is closer to $p_{\mathrm{mid}}'' = 1$ than in the other two panels.  Quantum dots have the highest communication rates, which is a direct result of our model using a fast clock rate and high collection efficiency.

%\begin{table}
%\label{tbl::optical_parameters}
%\centering
%\caption{Parameters for Optical Network Components}
%\begin{tabular}{m{3cm} m{4cm} m{1.3cm}}
%\textbf{Component} & \textbf{Parameter} & \textbf{Value}  \\ \hline
%\multirow{3}{*}{\parbox{3cm}{Single-photon detector (SNSPD)}} & Quantum efficiency 	& 0.70 \\
%										& Timing jitter 		& 10~ps \\
%										& Dark count rate	 	& 10~cps \\ \hline
%\multirow{4}{*}{\parbox{3cm}{Entangled-photon source (SPDC)}} & Pump period 			& 10~ns \\
%										& Generation probability & 0.02 \\
%										& Multi-pair probability & 2e-4 \\  \hline
%\end{tabular}
%\end{table}

\section{Discussion}
\label{sec::conclusion}
The key contributions of this paper are twofold.  First, we provide detailed instructions for the time-dependent operation of multiple quantum-networking protocols in one paper.  Second, we simulate the performance of networks using the quantum communication protocols.  We start with simulations using optimistic parameters, to compare how the protocols might perform on mature repeater technology.  We then simulate the protocols on multiple hardware platforms by selecting realistic performance parameters that are consistent with recent experimental demonstrations.  Taken together, this paper can be used as an engineering assessment for designing quantum networks and setting application-motivated milestones for the development of quantum hardware.  

This paper examined three protocols for creating distributed entanglement in a quantum repeater network.  The performance of these protocols in terms of communication rate was examined both analytically and with numerical simulation.  The different protocols offer complementary strategies for developing quantum networks.  The simplest protocol, \texttt{MeetInTheMiddle}, works best when signal transmission probability is relatively high.  Alternatively, the more complex \texttt{MidpointSource} can compensate for low signal transmission with a more sophisticated protocol and faster clocking.  Our simulations show that even hardware based on recent experimental demonstrations could demonstrate an advantage for \texttt{MidpointSource}, such as quantum dots operated at a fast clock rate.  A further advantage of \texttt{MidpointSource} is that the Bell-state measurement procedure is local to both repeaters sharing a link, allowing local tracking of clock-synchronization information~\cite{Kim2015}.

To see why we chose to implement two-detection schemes with loss heralding, one should consider the relative merits of our approach and its alternatives.  For example, other single-photon schemes exist that encode a qubit in the presence or absence of a photon~\cite{Duan2001,Simon2007,Sangouard2011}.  In this case, a single detection heralds entanglement, but there are two significant problems. First, the single-detection BSA cannot distinguish between one photon sent by one repeater and two photons sent (one from each repeater) where one is lost in transmission.  The probability for each repeater to emit a photon must be sufficiently small to suppress the likelihood of the double-emission event~\cite{Sangouard2011}.  Second, single-detection schemes are very sensitive to path-length fluctuations, which is problematic for long-distance fiber transmission~\cite{Duan2001,Simon2003,Chen2007}.  

While two-photon detection schemes address problems with single-photon detection, another concern is that loss heralding requires two-way communication with delays to confirm entanglement.  Relatively recent proposals consider only one-way communication with error correction to overcome the effects of loss~\cite{Fowler2010,Munro2012,Muralidharan2014,Ewert2015}.  One-way communication avoids the round-trip signaling delays, removing the need for long-lived quantum memory.  However, these proposals are very sensitive to photon loss for two reasons.  First, one-way protocols require much more sophisticated hardware, because the error correction requires many-qubit entangled states stored in quantum memory, the optical channel, or both.  The complexity overhead increases significantly with loss probability~\cite{Munro2012,Muralidharan2014}.  Second, Bennett~\emph{et al.} showed that one-way communication is impossible (\emph{i.e.} information capacity of the quantum channel is zero) if probability of qubit loss is 50\% or greater~\cite{Bennett1997_capacity}.  This rather general bound refers to the total loss during transmission between two repeaters, and it places a upper bound on inter-repeater distance if one transmits qubits through standard optical fiber~\cite{Fowler2010,Munro2012,Muralidharan2014}.

Compared to alternatives, two-way protocols with loss heralding require less device complexity and can function even in settings with greater than 50\% loss probability, making them suitable for near-term quantum repeater technology.  The one-way protocols may prove to have better network performance at later stages of technology maturation, when repeaters that operate on hundreds of qubits are achievable.  We argue that the designs considered here would use essentially the same hardware technology as one-way communication protocols, so that developing repeaters with loss-heralding protocols is a precursor to implementing one-way protocols.  In this way, developing repeaters based on loss-heralded protocols is prudent for near-term technology development. 

\begin{acknowledgments}
We thank Jim Harrington for suggesting improvements to the manuscript.
\end{acknowledgments}

\bibliography{repeaters}

\end{document}